\documentclass[twocolumn,showpacs,floatfix,superscriptaddress,amsmath,amssymb,aps]{revtex4}
\usepackage{mathrsfs}
\usepackage{amssymb}
\usepackage{graphicx}
\usepackage{hyperref}
\usepackage{ulem}
\usepackage{overpic}
\usepackage{color}
\usepackage{psfrag}
\usepackage{tabularx}
\usepackage{array}
\usepackage{placeins}
\makeatletter

\begin{document}
	
\title{Wigner-Yanase skew information, quantum entanglement and spin nematic quantum phase transitions in biquadratic spin-$1$ and spin-$2$ XY chains with single-ion anisotropies}

\author{Yan-Wei Dai}
\affiliation{Centre for Modern Physics,Chongqing University, Chongqing 400044, China}
\affiliation{Department of Physics, Chongqing University, Chongqing 400044, China}

\author{Sheng-Hao Li}
\affiliation{Chongqing Vocational Institute of Engineering, Chongqing 402260, China}

\author{Sam Young Cho}
\altaffiliation{E-mail: sycho@cqu.edu.cn}
\affiliation{Centre for Modern Physics,Chongqing University, Chongqing 400044, China}
\affiliation{Department of Physics, Chongqing University, Chongqing 400044, China}

\author{Huan-Qiang Zhou}
\affiliation{Centre for Modern Physics,Chongqing University, Chongqing 400044, China}

\begin{abstract}
 Quantum phase transitions between quantum uniaxial or biaxial spin nematic phases
 are investigated in biquadratic spin-$1$ and spin-$2$ XY infinite chains with the rhombic- and uniaxial-type single-ion anisotropies.
 Systematic discussions of distinctive singular behaviors are made to classify various types of quantum phase transition from one spin nematic state to the other spin nematic state in using the Wigner-Yanase skew information, the bipartite entanglement entropy, and the quadrupole moments.
 For the spin-$1$ system with the three uniaxial spin nematic quadrupole phases,
 we find that a discontinuous quantum phase transition, signaled by discontinuous behaviors of all the considered Wigner-Yanase skew information, bipartite quantum entanglement, and quadrupole moments, occurs from the $z$-ferroquadrupole phases to
 the $x$- or $y$-ferroquadrupole phases, while a continuous quantum phase transition occurs
 between the $x$- and $y$-ferroquadrupole phases.
 The central charge in the continuous phase transition line is estimated as
 $c \simeq 1$ from the entanglement entropy.
 Compared to the spin-$1$ system, depending on a given strength of the uniaxial-type single-ion anisotropy, the spin-$2$ system undergoes four different types of quantum phase transitions between the two biaxial spin nematic phases as the rhombic-type single-ion anisotropy varies:
 the quantum crossovers, connecting the two orthogonal biaxial spin nematic states adiabatically
 without an explicit phase transition, the continuous and the discontinuous phase transitions,
 and the spin nematic to magnetic transitions via the antiferromagnetic phase.
 In a sharp contrast to the spin-$1$ system, for the transitions between the two biaxial spin nematic phases, the discontinuous transition line is classified as a topological phase characterized by a doubly degenerate entanglement spectrum
 and a string order parameter defined by the Cartan generator of the $\mathrm{SO}(5)$ symmetry group in spin-$2$ systems,
 while the continuous phase transition is advocated by the central charge $c \simeq 1$.
 Whereas the continuous phase transition lines with $c \simeq 1/2$
 indicate that the transition between the biaxial spin nematic phase and the antiferromagnetic phase belongs to the Ising universality class.

\end{abstract}

\maketitle
	
	%%%%%%%%%%%%%%%%%%%%%%%%%%%%%%%%%%%
\section{Introduction}
 \label{section1}

 Quantum many-body interacting systems can order into different quantum states.
 As well as the quantum states themselves, quantum phase transition between quantum phases denoted by respective corresponding characteristic orders is one of the most important key phenomena  in understanding condensed matter physics \cite{Sachdev,Vojta,Book_Schollwock}.
 Novel quantum spin states arising from various quantum spin systems and quantum phase transitions revealed between these quantum spin states have received a lot of attention \cite{Book_Schollwock,Lacroix,Diep,Udagawa}.
 Of intriguing states are phases of interacting spins that
 do not order, making it impossible to characterize them by a local magnetic dipole moment or order. Such quantum spin states include topological (Haldane) phase \cite{Haldane,Pollmann12,Tonegawa,Okamoto,Tzeng12,Kjall}, valence bond solid \cite{Affleck1987, Schollwock1996, Balents10}, dimerized phase \cite{Barber,Klumper,Chubukov91,Solyom95}, quantum spin liquids \cite{Lacroix, Diep, Udagawa, Wen2019, Broholm,Savary2016,Knolle2019}, and spin nematic (quadrupolar) phase \cite{Blume1969, Matveev1973, Papanicolaou1988,Tsunetsugu2006, Lauchli2006, Penc2011,Andreev1984,Sudan,Zhitomirsky,Balents,Shannon,Ueda}.  Spin nematic states, as a fascinating example, have no local magnetic dipole moments, i.e.,
 no long-range magnetic dipole order but can be characterized by spin quadrupole moments forming a orientational order with breaking spin-rotational symmetry,
 which makes their experimental detection challenging with conventional experiments
\cite{Penc2011,Smerald2015,Smerald2016,Kohama2019}.
 Accordingly, characterizing such nonmagnetic spin states becomes a crucial and interesting problem in deeper understanding the exotic quantum phenomena of quantum spin systems.

 Fundamentally, quantum phase transition is a qualitative change in an intrinsic structure of ground state by quantum fluctuations with variations of system parameter. Naturally, there can be no universal order that is defined by an observable and characterizes various quantum many-body states. However, such intrinsic structural change in ground state can be revealed as a change in other aspects of the fundamental features of quantum many-body state, such as quantum correlations and quantum coherence as one of quantum mechanical fundamental properties. Even if it is not known a priori whether there is a quantum phase transition or what kind of phase there is at all, the quantumness of  many-body states can be used to identify quantum phase transitions.
 Indeed, quantum information measures such as, for instances, quantum entanglement measures \cite{Skrovseth,Vidal03,Osborne,Cho,Chung,Osterloh,Amico}, quantum mutual information \cite{Groisman,Adami,Alcaraz,Schumacher,Dai1,Dai2}, quantum coherence measures \cite{Baumgratz,Radhakrishnan,Mao}, and Wigner-Yaneses skew information (WYSI) \cite{WYSI,Wehrl,Streltsov,Luo03,Luo05,Karpat,Malvezzi,Li16,Lei,Qiu,Girolami13,Korzekwa,Yi19,Lin}  have been proposed to use as an indicator for quantum phase transitions and have been shown to capture them for quantum many-body systems.

 Such quantum information theoretical measures can provide the distinctive tools for the quest especially for novel types of quantum spin nematic states and quantum phase transitions between quantum spin nematic states.
 Thus, it would be a much more interesting problem to detect and identify different types of quantum phase transitions between nonmagnetic spin states without explicit magnetic ordering \cite{Book_Schollwock,Lacroix,Diep,Udagawa} such as, for instance, two orthogonal spin nematic states.
 In this study, we investigate the characteristic behaviors of the quantum information theoretical measures such as the bipartite quantum entanglement entropy and the WYSI, which received relatively less attention,
 for spin nematic quantum phase transitions.
% in biquadratic spin-$1$ and spin-$2$ XY chains with single-ion anisotropies.
 The quadrupole moments are investigated in connection with the characteristic behaviors of the entanglement entropy and the WYSI for the spin nematic quantum phase transitions.

 One-dimensional infinite spin-$1$ and spin-$2$ biquadratic XY models with
 the rhombic- and the uniaxial-type single-ion anisotropies are introduced
 in order to investigate spin nematic phases and quantum phase transitions between spin nematic phases.
 The one-dimensional infinite spin lattice systems are represented by using
 the infinite matrix product state (iMPS) representation and
 the ground sates are calculated with
 the infinite time-evolving block decimation (iTEBD) method \cite{Vidal03,Su12}.
 Our numerical results show that the bipartite entanglement entropy and the WYSI
 capture spin nematic phase boundaries consistent with each other
 for the spin-$1$ and -$2$ systems, respectively.
 However, the spin-$1$ and spin-$2$ phase diagrams are shown to have fundamental differences from each other although the spin-$1$ and -$2$ systems have the same form of the Hamiltonian.
 Connections between spin nematic phase transitions and
 tools of quantum information theory are studied for the spin-$1$ and -$2$ models by calculating the bipartite entanglement entropy including entanglement spectrum and the WYSI.
 According to a given uniaxial-type single-ion anisotropy, as the rhombic-type single-ion anisotropy varies, the bipartite entanglement entropy and WYSI show their characteristic features that are two discontinuities or one cusp for the spin-$1$ system,
 and gentle variation without inflection, one cusp or two cusps for the spin-$2$ system.
 The gentle variation without inflection of the spin-$2$ system implies  quantum crossover, i.e., the two biaxial spin nematic states are connected adiabatically without explicit phase transition at a specific anisotropic parameter.
 The cusp feature indicates an occurrence of continuous phase transition. However, for the one cusp behavior of the entanglement entropy and WYSI in the spin-$2$ system, two different type phase transitions are possible, i.e., discontinuous or continuous phase transitions can occur. If the cusp becomes shaper and diverges  as the truncation dimension increases, a continuous phase transition occur. Otherwise, i.e., if the cusp behavior doesn't change much and converges as the truncation dimension increases, the spin-$2$ system undergoes a discontinuous phase transition.
 Such characteristic behaviors of the bipartite entanglement entropy and the WYSI are discussed in connection with the corresponding behaviors of quadrupole order parameter.

 This paper is organized as follows.
 In Sec. \ref{section2}, the infinite biquadratic spin-$1$ and spin-$2$ XY chains with rhombic- and uniaxial-type single-ion anisotropies
 is introduced. The iMPS approach with the iTEBD
 is briefly explained in calculating ground state wavefunctions for the infinite chain models.
 The spin-$1$ and -$2$ nematic phase diagrams are briefly summarized.
 To clarify the relationship between quantum phase transitions and entanglement entropy and WYSIs, Sec. \ref{section3} devotes to discuss systematically the detailed behaviors of the entanglement entropy and the WYSIs
 for various types of quantum phase transitions between the uniaxial spin nematic phases
 for the spin-$1$ system or the biaxial spin nematic phases for the spin-$2$ systems.
 In Sec. \ref{section4}, the local magnetization and
 the local quadrupole order parameters are calculated and discussed
 the quantum spin nematic phases and the quantum spin nematic phase transitions between the uniaxial or the biaxial spin nematic phases in association with the entanglement entropy and
 the WYSIs. The consistent features depending on the transition types
 are given by discussing the detailed behaviors of the quadrupole order parameter
 for the uniaxial or the biaxial spin nematic states.
 A summary and remarks of this work is given in Sec. \ref{summary}.

%%%%%%%%%%%%%%%%%%%%%%%%%%%%%%%%%%
\section{Models and the spin nematic phase diagrams}
 \label{section2}
  We consider the biquadratic spin-$1$ and spin-$2$ XY models
  with the rhombic-type and uniaxial-type single-ion anisotropies.
  The biquadratic XY model Hamiltonian $H$ can be written as
\begin{equation}
 H = H_{BXY} + H_{R} + H_{D}
 \label{ham1}
\end{equation}
 with
\begin{eqnarray*}
 H_{BXY} &=& -J \sum_{j=-\infty}^\infty
 \left( S^{x}_jS^{x}_{j+1}+S^{y}_jS^{y}_{j+1}\right)^2, \\
 H_R &=& R \sum_{j=-\infty}^\infty [(S^{x}_j)^2-(S^{y}_j)^2], \\
  H_D &=& D \sum_{j=-\infty}^\infty  (S^{z}_j)^2,
\end{eqnarray*}
 where the biquadratic exchange interaction is denoted by $J(>0)$,
 and the rhombic- and the uniaxial-type single-ion anisotropies
 are denoted by $R$ and $D$, respectively.
 The same form of the Hamiltonian (\ref{ham1})
 is considered for spin-$1$ and spin-$2$ systems in order to compare the
 spin-$1$ and spin-$2$ spin nematic systems as one of purposes of our study.
 The spin-$1$ or spin-$2$ operators at site $j$ are denoted
 as $S^{\alpha}_{j}$ with $\alpha \in \{x, y, z\}$.
 It is known that a crystal field anisotropy and spin-orbit coupling may induce
 a microscopic effective Hamiltonian with single-ion anisotropies.
 Recently, such rhombic single-ion anisotropy effects have been
 investigated in the spin-$1$ Heisenberg model \cite{Tzeng}, XXZ model \cite{Ren},
 and biquadratic spin-$1$ and spin-$2$ XY models ($D=0$) \cite{Mao}.

 The one-dimensional infinite spin lattices of our model Hamiltonians are numerically
 studied with a wave function $\left|\psi\right\rangle$ of the Hamiltonian (\ref{ham1})
 in the iMPS representation.
 In order to calculate the numerical ground state $\left|\psi_G\right\rangle$,
 we have employed the iTEBD method
 \cite{Vidal03,Su12,Su13,Wang13,Dai17}.
 During the iTEBD procedure, the chosen initial state approaches to
 a ground state with the time step decreased from $dt = 0.1$ to $dt = 10^{-6}$
 according to a power law.
 A ground state wavefunction $|\psi_G\rangle$ in the iMPS representation
 for a given truncation dimension $\chi$ is yielded when the ground state energy
 is converged.
 The full ground state density matrix $\varrho_G = |\psi_G\rangle\langle\psi_G|$
 gives any reduced density matrix.
 For example, for a spin lattice-block (LB), the reduced density matrix $\varrho_{LB}$
 is given by tracing out the degrees of freedom of the rest of the
 LB, i.e., $\varrho_{LB} = \mathrm{Tr}_{(LB)^c} \, \varrho_G$
 with the full description of the ground state $|\psi_G\rangle$ in a pure state.

%%%%%%%%%%%%%%%%%%%%%%Fig. 1%%%%%%%%%%%%
    \begin{figure}
    \includegraphics[width=0.3\textwidth]{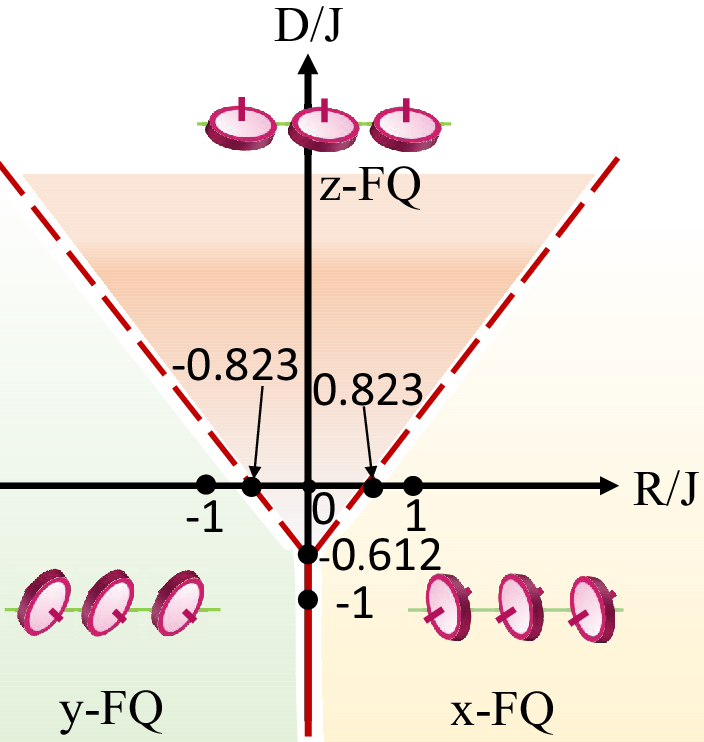}
    \caption{(color online) Ground state phase diagram for the biquadratic spin-$1$ XY
    model with the rhombic-type $(R)$ and the uniaxial-type $(D)$ single-ion anisotropies.
    The biquadratic spin exchange interaction is denoted by $J$.
    The ferroquadrupole phase denoted by $\alpha$-$FQ$ has the zero local spin fluctuation
    in the $\alpha$ axis $(\alpha \in \{x, y, z\})$.
    The local spin fluctuation in the plane perpendicular to the $\alpha$ axis at site $j$
    is denoted by a disk. Discontinuous phase transitions occur across the two dashed
    lines, while continuous phase transitions occur across the red solid line with
    the central charge $c \simeq 1$ on the $D/J$ axis.
    The multicritical point is located at $(R_M,D_M)=(0,0.612J)$. The detailed discussions are in the text.}
     \label{Fig1}
     \end{figure}
%%%%%%%%%%%%%%%%%%%%%Fig. 1%%%%%%%%%%%%

 %%%%%%%%%%%%%%%%%%Fig. 2%%%%%%%%%%%
    \begin{figure}
    \includegraphics[width=0.39\textwidth]{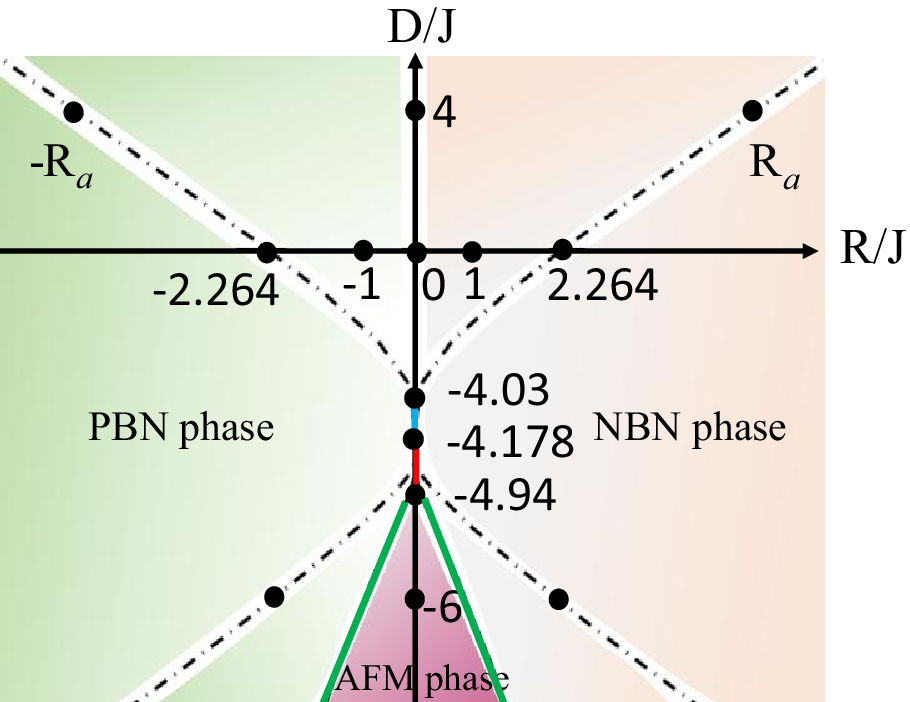}
    \caption{(color online) Ground state phase diagram for the biquadratic spin-$2$ XY
    model with the rhombic-type $(R)$ and the uniaxial-type $(D)$ single-ion anisotropies.
    The biquadratic spin exchange interaction is denoted by $J$.
    The spin nematic phases are determined by the quadrapole order parameter
    as the positive (negative) biaxial spin nematic (PBN (NBN)) phase for $\langle Q^{x^2-y^2} \rangle > 0$ $(<0)$. 
    The four dash-dot lines and the $R=0$ line indicate the uniaxial nematic states inside the positive or negative biaxial spin nematic phase. 
    Continuous phase transitions between the antiferromagnetic (AFM) phase and
    PBN or NBN phases occur across the green solid lines with the central charge $c \simeq 1/2$.
    The red solid line on the $D/J$ axis indicates the topological phase characterized by
    a string order parameter defined by the Cartan generator $L^{12}$ of the SO(5) symmetry
    group in spin-$2$ systems.
    The critical end-point (CEP) with $ \langle Q^{\alpha\alpha}\rangle =0$ $(\alpha \in \{x, y, z\})$ is located at $(R_{CEP}, D_{CEP}) = (0, -4.03J)$.
    The blue solid line $D(-4.178J,-4.03J)$ on the $D/J$ axis indicates a critical phase with
    the central charge $c \simeq 1$.
    The detailed discussion is in the text.
    }
    \label{Fig2}
    \end{figure}
%%%%%%%%%%%%%%%%%%Fig.2%%%%%%%%%%%%%%%%%%
% In the absence of the uniaxial-type single-ion anisotropy, i.e., $D=0$,
% the three uniaxial spin nematic quadrupole phases for the spin-$1$ system and the
% two biaxial spin nematic phases for the spin-$2$ system were identified in Ref.
% \onlinecite{Mao}.
% In this study, for $D \neq 0$, we have interested in what kind of quantum phase transitions are possible between spin nematic phases for each spin system.
 From the iMPS ground state $|\psi_G\rangle$, we have studied
 the bipartite entanglement entropy, the WYSI, and the quadruple order parameters.
 As a brief summary, we draw the schematic spin nematic phase diagrams
 in the parameter space $(R/J,D/J)$ with $J > 0$ for the truncation dimension $\chi = 150$.
 The spin nematic phase diagrams for the spin-$1$ and spin-$2$ systems
 are displayed in Fig.~\ref{Fig1} and Fig.~\ref{Fig2}, respectively.
 The spin-$1$ nematic phase diagram in Fig. \ref{Fig1} shows
 that the whole parameter space $(R/J,D/J)$ is divided into the three regions by the three distinct ferroquadrupole phases. Direct continuous quantum phase transitions with the central charge $c \simeq 1$
 occur between the $x$- and the $y$-ferroquadrupole phases along the $R=0$ solid line
 for $D < -0.612J$,
 while direct discontinuous quantum phase transitions occur between the $z$-ferroquadrupole
 phase and $x$-ferroquadrupole phase or between the $z$-ferroquadrupole
 phase and $y$-ferroquadrupole phase along the two dashed straight lines merging
 at the multicritical point $(R_M,D_M)=(0,-0.612J)$.

 Compared to the spin-$1$ system,
 the spin-$2$ nematic phase diagram in Fig. \ref{Fig2} shows
 that there are the two distinct biaxial spin nematic phases. The spin-$2$
 antiferromagnetic (AFM) phase appears for the negative uniaxial-type anisotropy relatively
 much larger than the rhombic-type anisotropy.
 Direct continuous or discontinuous phase transitions occur
 between the two biaxial spin nematic phases across the critical and the topological phases
 on the $R=0$ line for $-4.178J < R < -4.03J$ (blue solid line) and $-4.94J < R < -4.178J$ (red solid line), respectively, as well as the quantum crossover
 with the crossover range between the uniaxial points, $\pm R_a$, for $D > -4.03 J$.
 In the $(R,D)$-parameter space, the critical end-point and the multicritical point are located at $(R_{CEP},D_{CEP}) = (0,-4.03J)$ and $(R_M,D_M) = (0,-4.94J)$, respectively.
 A transition can also occur between the two biaxial spin nematic phases via the antiferromagnetic phase for $R < -4.94J$, as the rhombic-type single-ion anisotropy varies.
 The continuous phase transition between a biaxial spin nematic phase and the antiferromagnetic phase belongs to the Ising universality class with the central charge $c = 1/2$.

%%%%%%%%%%%%%%%%%%%%%%%%%%%%%%%%%%%%%%%%%%%%%%%%
%%%%%%%%%%%%%%%%%%%%%%%%%%%%%%%%%%%%%%%%%%%%%%%%

\section{Wigner-Yanase skew information, entanglement entropy, and quantum phase transitions}
 \label{section3}
 A most unique quantum feature of composite systems is quantum entanglement.
 Its intrinsic properties are studied and used in various research areas such as
 quantum technology and quantum information science \cite{Nielsen,Horodecki09},
 and have potential applications such as quantum teleportation \cite{Bouwmeester},
 quantum computation \cite{Horodecki09,Jozsa} and quantum sensing \cite{Degen}.
 Also quantum entanglement is applicable to investigate quantum many-body systems, e.g.,
 especially for quantum phase transitions \cite{Osterloh,Eisert,Vidal03,Calabrese,Amico} because
 it exhibits a characteristic singular behavior, i.e., nonanalyticity
 \cite{Wu04,Fradkin06}
  at critical points of the quantum many-body systems. Various entanglement measures have been recently demonstrated to be a crucial tool
 in detecting, classifying, and understanding quantum phase transitions \cite{Vidal03,Calabrese,Amico}.
 As one of the widely used quantitative entanglement measures, entanglement entropy
 is defined through the von Neumann entropy and
 its scaling behavior of entanglement entropy with a logarithmic divergence \cite{Vidal03,Calabrese}
 can be particularly useful to characterize near a quantum phase transition.

 To investigate the transition between the spin nematic phases,
 we consider the entanglement entropy
 for a bipartite spin chain partitioned into two subsystems, i.e.,
 the left semi-infinite spin chain $L(-\infty, \cdots, j-1)$ and
 the right semi-infinite spin chain $R(j, \cdots, \infty)$.
 In terms of the left (right) Schmidt basis $|\psi^L_\alpha\rangle (|\psi^R_\alpha\rangle)$for the left (right) semi-infinite spin chain
 $L(R)$, an iMPS ground state wave function can be expressed with their Schmidt coefficient $\lambda_\alpha$ as
\begin{equation}
 |\psi_G \rangle = \sum_{\alpha=1}^\chi \lambda_\alpha
  |\psi^L_\alpha\rangle |\psi^R_\alpha\rangle,
\end{equation}
 where $\chi$ is the truncation dimension in the iMPS representation.
 In the iMPS representation,
 the entanglement entropy of the ground state $|\psi_G\rangle$ is thus defined as the von Neumann entropy of the reduced density matrix $\varrho_L$ or $\varrho_R$.
 Explicitly, the entanglement entropy is given as
 $S = -\mathrm{Tr} \, \varrho_{L} \log_2 \varrho_{L}=-\mathrm{Tr} \, \varrho_{R} \log_2 \varrho_{R}$.
 In terms of the Schmidt decomposition coefficients $\lambda_\alpha$ in the iMPS representation,
 the entanglement entropy $S$ can be
 written as~\cite{Tagliacozzo,Pollmann0}:
\begin{equation}
 S = -\sum_{\alpha=1}^\chi \lambda_\alpha^2 \log_2 \lambda_\alpha^2.
 \label{SS}
\end{equation}

 Together with the entanglement entropy,
 we study the behaviors of the WYSI for quantum phase transitions between the spin nematic phases.
 As a measure of the noncommutativity of a quantum state $\varrho$ and an observable $K$,
 which is called skew information,
 the WYSI \cite{WYSI,Wehrl,Streltsov}
 is defined as
\begin{equation}
   I(\varrho, K) = - \frac{1}{2}\mathrm{ Tr}[\sqrt{\varrho}, K]^2,
 \label{SYWI}
\end{equation}
 where $K$ is a nondegenerate Hermitian matrix and
  $[\sqrt{\varrho}, K] = \sqrt{\varrho} K - K \sqrt{\varrho}$.
 The information content of a quantum state $\varrho$ with
 respect to an observable $K$ can be quantified by the WYSI.
 Then, the WYSI reflects the information of the quantum state $\varrho$
 skewed to the observable $K$ \cite{WYSI} or
 the quantum uncertainty of the observable in the quantum state \cite{Luo03,Luo05}.

 For a bipartite composite state $\varrho_{AB}$
 with the two subsystems $A$ and $B$, the local
 quantum coherence or asymmetry with respect to the first subsystem
 can be defined as $I(\varrho_{AB}, K_A \otimes \mathbb{I}_B)$ with
 the unit matrix $\mathbb{I}$.
 The skew information $I(\varrho_{AB},K_A \otimes \mathbb{I}_B)$
 is related to the uncertainty of measuring
 the observable $K_A$ with respect to the composite state $\varrho_{AB}$.
 Pure quantum states $\varrho_{p}$ have the reduced form of the WYSI
 as the variance $I(\varrho_p ,K)=V (\varrho_p,K)
 = \mathrm{ Tr} \varrho_p K^2 -  \left(\mathrm{ Tr}\varrho_p K \right)^2$
 of the observable $K$ in the state $\varrho$ \cite{Luo03}.
 For our spin-$1$ and spin-$2$ chains, we thus the nearest-neighbor two spins,
 i.e., $S_j$ and $S_{j+1}$, in order to calculate the WYSI in Eq. (\ref{SYWI}).
 For the two spins,
 the quantum state $\varrho$ becomes the two-site reduced density matrix
 $\varrho=\varrho_{j,j+1}=\mathrm{Tr}_{L^c} \varrho_G$
 with $L=(j,j+1)$ and the observable $K$ can be chosen as the spin operator $S^{\alpha}
 (\alpha\in\{x, y, z\})$, i.e., $K^\alpha = K_j^{\alpha}\otimes \mathbb{I}_{j+1}$.
 The local skew information $I^\alpha = I(\varrho, K^\alpha)$ defined as
\begin{equation}
 I(\varrho, K^\alpha) = I(\varrho_{j,j+1},K^\alpha_j \otimes \mathbb{I}_{j+1})
 \label{eq3}
\end{equation}
 is calculated to investigate the characteristic properties for the spin nematic states.
 Due to the fact that the systems we consider are invariant upon
 exchanging the two spins, the local WYSI $I(\varrho_{j,j+1},K^\alpha_j \otimes
 \mathbb{I}_{j+1})$ remains also unchanged.

%%%%%%%%%%%%%%%%%%%%%%%%%%%%%%%%%%%%%%%%%%%
%
\subsection{Quantum phase transitions for spin-$1$ system}
\label{spin-1-system}
 As show in the spin nematic phase diagram in Fig. \ref{Fig1},
 the spin-$1$ XY models undergo the two types of phase transitions between the three ferroquadrupole phases on the anisotropy parameter space $(R,D)$.
 One is the discontinuous phase transition, the other is the continuous phase transition.
 The phase boundaries for the continuous or discontinues phase transitions
 can be determined by understanding the characteristic behaviors of the entanglement entropy and the WYSI.
 We will discuss the detailed characteristic behaviors of the entanglement entropy and the WYSI as the rhombic-type anisotropy $R$ varies
 for a chosen value of the uniaxial-type anisotropy $D$.

%%%%%%%%%%%%%%%%%%%%%%%%%%%%%%Fig3%%%%%%%%%%%%%%%%%%%%%
 \begin{figure}
    \includegraphics[width=0.4\textwidth]{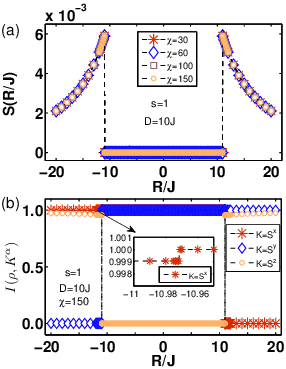}
    \caption{(color online) (a) Entanglement entropy $S$ for various truncation dimensions $\chi=30, 60, 100$ and $150$, and (b) WYSI $I^\alpha=I(\varrho, K^{\alpha})$
    for $\chi=150$ in the spin-$1$ system with the uniaxial-type anisotropy $D=10J$.
    Here, $K=S^{\alpha}$ ($\alpha\in\{x, y, z\}$) with the spin-$1$ operator $S^{\alpha}$.
    In the inset of (b), $I^x$ is plotted near the jump at $R = -10.972J$.
    $I^y$ has a similar jump at $R= 10.972J$, which is not presented here.
    } \label{Fig3}
     \end{figure}
%%%%%%%%%%%%%%%%%%%%%%%%%%%%Fig3%%%%%%%%%%%%%%%%%%%%%%%%%

 \subsubsection{Discontinuous spin nematic to nematic quantum phase transitions}
 \label{dis-spin1}
 The two-spin reduced density matrix $\varrho=\varrho_{j,j+1}$, obtained
 from the iMPS ground state $|\psi_G\rangle$, gives the bipartite entanglement entropy $S$ and the skew information $I^\alpha=I(\varrho,K^\alpha)$.
 In Fig. \ref{Fig3}, we plot (a) the $S$ and
 (b) the WYSIs $I^\alpha$  as a function of $R/J$ for $D = 10 J$.
 Figure \ref{Fig3} (a) shows clearly
 that for the truncation dimensions from $\chi=30$ to $\chi=150$,
 the entanglement entropy $S$ are discontinuous at $R =\pm 10.972 J$.
 These two nonanalyticities of the entanglement entropy indicate
 the occurrences of the discontinuous quantum phase transitions
 from the $z$-ferroquadrupole phase to the $y$-ferroquadrupole phase at $R = - 10.972 J$
 or the $x$-ferroquadrupole phases at $R = + 10.972 J \equiv R_c$, respectively.
 Note that $S=0$ for the $z$-ferroduadrupole phase, which implies
 that the ground state is in a product state for $-R_c < R < R_c$.
 It is also shown clearly that for $R > R_c$ and $R < -R_c$
 the amplitude of the entanglement does not change as the truncation dimension becomes bigger from $\chi=30$ to $\chi=150$, which implies that the spin-$1$ chain is noncritical.
 The entanglement entropy satisfies the relation ship $S(R)=S(-R)$ for $J > 0$.

 Similar to the entanglement entropy,
 Fig. \ref{Fig3}(b)
 shows that the WYSI $I^z$ undergoes two prominent jumps at $R = \pm R_c$, respectively,
 which implies that the $I^z$ captures the two discontinuous quantum phase transitions
 at $R = \pm R_c$,
 and $I^z(R) = I^z(-R)$.
 Also, $I^z =0$ for $-R_c < R < R_c$, i.e., the $z$-ferromagnetic phase and
 $I^z \simeq 1$ for $R > R_c$, i.e., the $y$-ferroquadrupole phase
 and $R < -R_c$, i.e., the $x$-ferroquadrupole phase.
 However, in contrast to the $I^z$,
 there seems to be an abrupt jump only at $R = R_c$ ($R =-R_c$) for $I^{x}$($I^y$)
 in Fig. \ref{Fig3}(b),
 and thus there seems to have the relationship $I^{x}(R) = I^{y}(-R)$
 rather than $I^{x/y}(R) = I^{x/y}(-R)$.
 However, if we take a closer look near the transition points $R = \pm R_c$,
 it is found that $I^{x}$ ($I^{y}$) undergoes a noticeable small jump at $R = -R_c$ ($R = R_c$)
 as shown in the inset of Fig. \ref{Fig3}(b).
 Consequently,
 the discontinuous behaviors, i.e., the nonanalyticity of the WYSIs $I^{x/y/z}$ at $R=\pm R_c$
 indicate the occurrence of the discontinuous quantum phase transitions.
 %

%%%%%%%%%%%%%%%%%%%%%%Fig.4%%%%%%%%%%%%%%%%%%%%%%%%%%%%%
 \begin{figure}
    \includegraphics[width=0.4\textwidth]{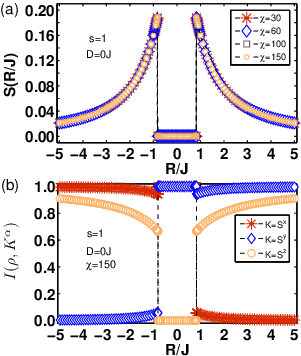}
    \caption{(color online) (a) Entanglement entropy $S$ for various truncation dimensions $\chi=30, 60, 100$ and $150$, and (b) WYSI $I^\alpha=I(\varrho, K^{\alpha})$
    for $\chi=150$ in the spin-$1$ system without the uniaxial-type anisotropy $D=0$.
    } \label{Fig4}
     \end{figure}
%%%%%%%%%%%%%%%%%%%%%Fig.4%%%%%%%%%%%%%%%%%%%%%%%%%%%%%%%%
%
 In the absence of the uniaxial-type anisotropy $D=0$,
 the entanglement entropy and the WYSIs are plotted in Fig. \ref{Fig4}.
 The $S$ and $I^{x/y/z}$ for $D = 0$ exhibit very similar behaviors
 with those for $D = 10J$, respectively, except for the transition points,
 i.e., $R=\pm 0.823 J$ for $D = 0$.
 The transition points captured by the $S$ and the WYSIs $I^{x/y/z}$
 are consistent with those detected from the quantum coherence measures, i.e., the $l_1$
 norm of coherence $C_{l_1}(\rho)$ \cite{Baumgratz}, the relative entropy of coherence
 $C_{re}(\rho)$ \cite{Baumgratz}, and the quantum Jensen-Shannon divergence $C_{JS}(\rho)$
 \cite{Radhakrishnan}, in the absence of the uniaxial anisotropy $D=0$ in Ref.
 \onlinecite{Mao}.
 Noticeably,
 at the transition points between the $z$-ferroquadrupole phase and the $x/y$-ferroqudarupole phases, the nonanalyticities of $I^{x/y}$ for $D=0$ in Fig. \ref{Fig4}(b) are clearer compared to those for $D=10J$ in Fig. \ref{Fig3}(b).
 Actually, the larger the $D$, the smaller the discontinuous gap of $I^x$($I^y$) at $R=
 R_c$ ($R=-R_c$).

 As shown in the phase diagram of Fig. \ref{Fig1},
 the value of $R_c$ depends on the uniaxial single-ion anisotropy $D$
 bigger than $D = -0.612J$ at the multicritical point.
 It seems that the stronger the positive anisotropy, the bigger the value of $R_c$ linearly.
 The two linear phase boundaries given by $R=\pm R_c$ clearly separate the three spin
 nematic phases, shown in the phase diagram in Fig. \ref{Fig1}.
 The phase boundaries are numerically fitted with the linear fitting function $D_c/J = \pm\, a\, R_c/J + b$ with the fitting coefficients $a = 0.98(2)$ and $b = -0.78(8)$. Thus, the estimate value of the multicritical point from the fitting
 is given as  $D_c =-0.78(8)$. The estimate value from the entanglement entropy $S$
 and the WYSIs $I^{x/y/z}$ is $ D_c = -0.612J$. The discrepancy between the two estimates
 are from the nonlinearity of the actual phase boundary near the multicritical point.

 Note that $I^x\simeq 0$ for $R \gg  R_c$, $I^y \simeq 0$ for $R \ll R_c$, and $I^z=0$
 for $-R_c < R < R_c$.
 The $x/y/z$-ferroquadrupole phase correspond to $I(\rho, K^{z/y/z})=0$, respectively, in the spin nematic phase diagram in Fig. \ref{Fig1}. The $I(\varrho,K^\alpha)=0$ implies that
 the $\alpha$-ferroquadrupole state $\varrho$ is left invariant by measuring the observable
 $K^\alpha$ \cite{Girolami13}.
 The skew information  $I(\varrho,K^\alpha)=0$ is related to the uncertainty of measuring
 the observable $S^\alpha_j$ with respect to the composite state $\varrho_{j,j+1}$.
 Consequently, such zero WYSIs mean that quantum uncertainty cannot be observed
 and classical ignorance gives rise to statistical errors because
 classical mixing does not influence on the corresponding WYSI \cite{Karpat,Korzekwa},
 i.e.,
 the WYSI is a mixture of the eigenstate of the observable in the eigenbasis of the
 $S^\alpha_j\otimes \mathbb{I}_{j+1}$.

%%%%%%%%%%%%%%%%%%%%%%%%%%%%%%%Fig5%%%%%%%%%%%%%%%%%%%%
 \begin{figure}
    \includegraphics[width=0.4\textwidth]{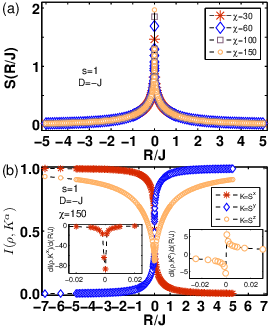}
    \caption{(color online) (a) Entanglement entropy $S$ for various truncation dimensions $\chi=30, 60, 100$ and $150$, and (b) WYSI $I^\alpha=I(\varrho, K^{\alpha})$
    for $\chi=150$ in the spin-$1$ system with the uniaxial-type anisotropy $D=-J$.
    The insets display the first-order derivatives of the WYSIs $I^{x}$ and $I^z$.
    } \label{Fig5}
     \end{figure}
%%%%%%%%%%%%%%%%%%%%%%%%%%%%Fig5%%%%%%%%%%%%%%%%%%%%%%%%%
 \subsubsection{Continuous spin nematic to nematic quantum phase transitions}
 \label{con-spin1}
 As was discussed, the two discontinuous phase transition boundaries meet at
 the multicritical point $(R_M,D_M)=(0, -0.612J)$.
 Thus, the uniaxial anisotropy $D = -J$ is chosen to discuss 
 the behaviors of the entanglement entropy and the WYSIs for $D < D_c = -0.612J$.
 In Fig.~\ref{Fig5}, we plot (a) the entanglement entropy $S$ and 
 (b) the $I(\varrho, K^{\alpha})$ as a function of $R/J$.
 In contrast to the cases of $D=10J$ and $D=0$, respectively shown in Figs. \ref{Fig3}(a) and \ref{Fig4}(a), Fig.~\ref{Fig5}(a) shows that
 the entanglement entropy is continuous for a given truncation dimension $\chi$ and its cusp appears at $R=0$.
 The cusp peak value at $R=0$ increases as the truncation dimension $\chi$ increases
 from $\chi=30$ to $\chi=150$.
 Actually, such an increment of entanglement according to an increment of the truncation dimension indicates the divergence of entanglement entropy in the thermodynamic limit $\chi \rightarrow \infty$, which means that the ground state is critical at $R=0$.
 This critical point will be classified by calculating the central charge in Sec. \ref{phase-rhom}.
 As a result, the implication of the cusp behavior of the entanglement entropy at $R=0$ is
 that the spin-$1$ chain system undergoes a continuous quantum phase transition
 between the $x$- and the $y$-ferroquadrupole phases.

 Figure~\ref{Fig5}(b) also shows that $I^x(R)=I^y(-R)$ as well as $I^z(R)=I^z(-R)$ but in contrast to the discontinuous phase transitions in Figs. \ref{Fig3}(b) and \ref{Fig4}(b),
 the WYSIs $I^{x/y/z}$ are continuous for the whole parameter range of the rhombic-type anisotropy $R$ for the uniaxial-type anisotropy $D=-J$.
 The $I^{x/y}$ have an inflection point at $R=0$, while the $I^z$ has a cusp at the same $R=0$.
 The behaviors of the WYSIs at $R=0$ indicate the nonanalyticity of the WYSIs at $R=0$.
 To be clear, we display the first-order derivatives of the WYSIs $I^x$ and $I^z$ in
 the inset of Fig. \ref{Fig5} (b). In fact, the derivative of the $I^y$ (not presented here)
 exhibits a similar behavior with those of the $I^x$.
 The cusp singular behaviors are revealed in the first-order derivatives of the $I^{x/y}$ at the inflection point at which a change in the direction of WYSI's curvature occurs.
 Also the derivative of the $I^z$ shows the discontinuous behavior at the cusp singular point. Such crucial behaviors of the skew informations $I^{x/y/z}$ and
 the derivatives indicate an occurrence of continuous quantum phase transition across $R=0$ \cite{Karpat,Yi19}.  Thus, occurring between the spin nematic phases, i.e., $x/y$-ferroquadrupole phases along the parameter line $R=0$ for $D < D_c$ in Fig. \ref{Fig1}, the continuous quantum phase transition is captured by the characteristic behaviors of the WYSIs $I^{x/y/z}$.

%%%%%%%%%%%%%%%%%%%%%%%%%%%%Fig.6%%%%%%%%%%%%%%%%%%%%%%%
 \begin{figure}
    \includegraphics[width=0.4\textwidth]{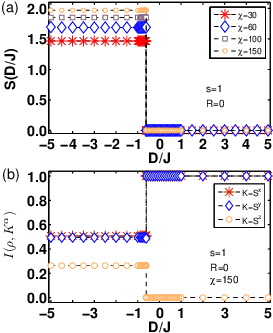}
    \caption{(color online) (a) Entanglement entropy $S$ for various truncation dimensions $\chi=30, 60, 100$ and $150$, and (b) WYSI $I^\alpha=I(\varrho, K^{\alpha})$
    for $\chi=150$ in the spin-$1$ system with the rhombic-type anisotropy $R=0$.
    } \label{Fig6}
     \end{figure}
%%%%%%%%%%%%%%%%%%%%%%%%%%%Fig.6%%%%%%%%%%%%%%%%%%%%%%%%%%
 \subsubsection{Phase diagram of the biquadratic spin-$1$ XY chain with the uniaxial-type single-ion anisotropy}
 \label{phase-rhom}
 In the absence of the rhombic-type anisotropy $R=0$,
 the model Hamiltonian in Eq. (\ref{ham1}) becomes
 the biquadratic spin-$1$ or spin-$2$ XY models with the uniaxial-type single-ion anisotropy
 $D$, i.e., the Hamiltonian (\ref{ham1}) reduces to
 \begin{equation}
 H_{BD} = -J \sum_{j=\infty}^\infty
 \left( S^{x}_jS^{x}_{j+1}+S^{y}_jS^{y}_{j+1}\right)^2
  + D \sum_{j=\infty}^\infty  (S^{z}_j)^2.
 \label{ham2}
\end{equation}
 In the limit of large positive anisotropy $D \gg J$, the Hamiltonian (\ref{ham2}) becomes
 $H_{BD} \approx \sum (S^z_j)^2$ and the system is in
 the $z$-ferroquadrupole phase.
 As we discussed in Sec. \ref{dis-spin1}, the $z$-ferroquadrupole phase has
 $I(\varrho,K^z)=0$ and $I(\varrho,K^{x/y})=1$, shown in Fig. \ref{Fig3}.

 We plot (a) the entanglement entropy $S$ and (b) the WYSIs as
 a function of the uniaxial-type anisotropy $D/J$ in Fig. \ref{Fig6}.
 Figure \ref{Fig6}(a) shows clearly
 that the entanglement entropies $S$ from $\chi=30$ to $\chi=150$
 are discontinuous at $D =-0.612 J \equiv D_c$.
 As the truncation dimension increases from $\chi=30$ to $\chi=150$,
 the entanglement entropy does not change, i.e., $S=0$
 in the $z$-ferroquadrupole phase for $D > D_c$,
 while it becomes bigger for $D < D_c$.
 As we discussed in Sec. \ref{con-spin1},
 the increment of the entanglement entropy according to an increment of the truncation dimension implies that the spin-$1$ chain system is in a critical phase for $D < D_c$.
 At $D=D_c$, the nonanalyticity of the entanglement entropy indicates
 the occurrence of the discontinuous quantum phase transition
 from the $z$-ferroquadrupole phase to the critical phase.
 Similar to the entanglement entropy,
 Fig. \ref{Fig6}(b) shows that the WYSIs $I^{x/y/z}$ are discontinuous and
 undergo abrupt jumps at $D = D_c$.
 As a result, the WYSIs $I(\varrho,K^{x/y/z})$
 is shown to capture the discontinuous phase transition between
 the critical phase for $D < D_c$ and the $z$-ferroquadrupole phase for $D > D_c$.

%%%%%%%%%%%%%%%%%%%%%%%Fig7%%%%%%%%%%%%%%%%%%%%%%%%
\begin{figure}
    \includegraphics[width=0.46\textwidth]{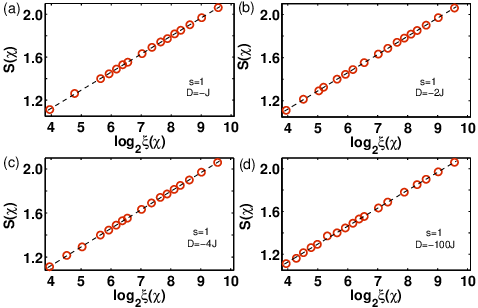}
    \caption{(color online) Entanglement entropy $S$ versus logarithmic correlation length $\log_2\xi$ for (a) $D=-J$, (b) $D =-2J$, (c) $D= -4J$ and (d) $D =-100J$ in the biquadratic spin-$1$ XY chain with the uniaxial-type anisotropy $D$ in Eq. (\ref{ham2}).
    The entanglement entropy and the correlation were obtained for truncation dimensions from $\chi=10$ to $\chi=200$ in the iMPS representation approach.
    The central charges are estimated as $c \simeq 1$ in the critical phase.
    } \label{Fig7}
     \end{figure}
%%%%%%%%%%%%%%%%%%%%%Fig7%%%%%%%%%%%%%%%%%%%%%%%%%%%%

%
 The characterization of the critical phase is possible by studying the scaling of entanglement
 entropy predicted by conformal field theory, i.e., the von Neumann entropy between the left and the right semi-infinite spin chains scales as
 \cite{Korepin0, Tagliacozzo, Pollmann0, Hu0,Calabrese}
 \begin{equation}
 S(\chi) = \frac{c}{6} \log_2 \xi(\chi) + S_0,
 \label{entropy1}
 \end{equation}
 where $S_0$ is a non-universal constant.
 The iMPS representation gives the correlation length
 $\xi(\chi)$ defined as $\xi(\chi)=1/\log_2(\varepsilon_1(\chi)/\varepsilon_2(\chi))$ with $\varepsilon_1(\chi)$ and $\varepsilon_2(\chi)$ being the largest and the second largest eigenvalues of the transfer matrix for a given truncation dimension $\chi$, respectively.
 Thus, the iMPS representation approach provide a way to characterize critical systems by using its scaling property of the entanglement entropy. In the critical phase ($R=0$ and $D < D_c$) of the spin-$1$ system, the scaling relation between the entanglement entropy $S(\chi)$ and the logarithmic correlation length $\log_2 \xi(\chi)$
 can be shown with various truncation dimensions
 from $\chi=10$ to $\chi=200$ for $D = -J$, $-2J$, $-4J$, and $-100J$ in Fig. \ref{Fig7}.
 According to the scaling relation in Eq. (\ref{entropy1}),
 one can estimate the central charges $c$ by performing the numerical fitting with the fitting function $S(\chi) = \frac{c}{6}  \log_2 \xi(\chi) + S_0$.
  The fitting coefficients are given as
  (a) $c/6=0.168(1)$, $S_0=0.452(8)$ for $D = -J$,
  (b) $c/6=0.1687(8)$, $S_0=0.448(5)$ for $D = -2 J$,
  (c) $c/6=0.169(1)$, $S_0=0.449(7)$ for $D = -4 J$ and
  (d) $c/6=0.168(2)$, $S_0=0.45(1)$ for $D = -100 J$.
 The estimate central charges are summarized in Table \ref{table1}.
 Consequently,
 in the absence of the rhombic-type single-ion anisotropy $R=0$, the critical phase
 for $D < D_c$ is characterized by the estimate central charge $c \simeq 1$.

%%%%%%%%%%%%%%%%%%%%%%%%%%%%TABLE 1%%%%%%%%%%%%%%%%%%%%%%%
\begin{table}
\renewcommand\arraystretch{2}
\caption{Estimate central charges $c$ for various values of the uniaxial single-ion anisotropy $D$
in the critical phase of the spin-$1$ system with the rhombic-type anisotropy $R=0$.}
\begin{tabular}{c|cccccc}
\hline
\hline
 $D$ & \begin{minipage}{1.4cm} $-J$  \end{minipage} &
  \begin{minipage}{1.4cm} $-2J$  \end{minipage} &
  \begin{minipage}{1.4cm} $-4J$  \end{minipage} &
  \begin{minipage}{1.4cm} $-100J$  \end{minipage} & \\
 \hline
 \begin{minipage}{1.2cm}  $c$ \end{minipage}
 &1.009(7) & 1.012(5) & 1.012(6) & 1.01(1)\\
 \hline
 \hline
\end{tabular}
\label{table1}
\end{table}
%%%%%%%%%%%%%%%%%%%%%%%%%%%%TABLE 1%%%%%%%%%%%%%%%%%%%%%%%

%%%%%%%%%%%%%%%%%%%%%%%%
%
\subsection{Qantum phase transitions for spin-$2$ system}
 Next, we consider the biquadratic spin-$2$ XY chain with the rhombic- and the
 uniaxial-type single-ion anisotropies. The spin-$2$ chain system has the same form with
 the spin-$1$ Hamiltonian (\ref{ham1}).
 Although the two spin-$1$ and -$2$ systems have the same form with the Hamiltonian
 (\ref{ham1}),
 their quantum phase diagrams are remarkably different each other, as shown in Figs.
 \ref{Fig1} and \ref{Fig2}, respectively.
 As summarized in Fig. \ref{Fig2}, the spin-$2$ system has the positive or negative
 biaxial spin nematic phases depending on the sign of the strength of the rhombic-type
 single-ion  anisotropy $R$, respectively.
 For the positive order parameter $\langle Q^{x^2-y^2}_i\rangle > 0$ without
 magnetic moments $\langle S^\alpha_i \rangle =0$, the system is in the positive biaxial spin nematic (PBSN) phase,
 while for the negative order parameter $\langle Q^{x^2-y^2}_i\rangle < 0$ without
 magnetic moments $\langle S^\alpha_i \rangle =0$,
 the system is in the negative biaxial spin nematic (NBSN) phase.
 It might be interesting to investigate how quantum spin nematic phase transition between the positive and the negative spin nematic phases of the spin-$2$ system occurs
 as the rhombic-type anisotropy $R$ varies for a given uniaxial-type single-ion anisotropy $D$.

%%%%%%%%%%%%%%%%%%%%%%%%%%%%%%%%%%%%%%%%%%%%%%%%%%%%%%%%%%%
\begin{figure}
    \includegraphics[width=0.4\textwidth]{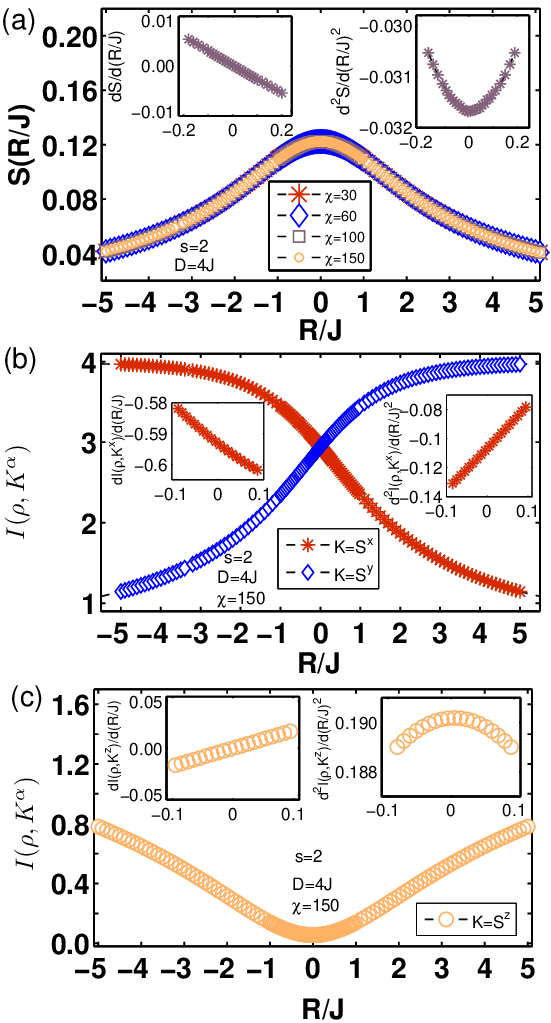}
    \caption{(color online) (a) Entanglement entropy $S$ for various truncation dimensions $\chi=30, 60, 100$ and $150$, (b) WYSI $I^x=I(\varrho, K^x)$ and $I^y=I(\varrho, K^y)$, and (c) $I^z=I(\varrho, K^{z})$
    for $\chi=150$ in the spin-$2$ system with the uniaxial-type anisotropy $D = 4J$.
    In the insets, the first- and second-order derivatives of $S$ and $I^{x/y/z}$
    are plotted.
    } \label{Fig8}
     \end{figure}
%%%%%%%%%%%%%%%%%%%%%%%%%%%%%%%%%%%%%%%%%%%%%%%%%%%%%%%%%%%

%
 \subsubsection{Quantum crossover between biaxial spin nematic phases}
 \label{qcrossover}
 In Fig. \ref{Fig8}, we plot (a) the entanglement entropy $S$,
 (b) the WYSIs $I^{x/y}$ and (c) $I^z$ as a function of $R/J$ for $D = 4J$.
 Figure \ref{Fig8}(a) shows that
 the entanglement entropy is continuous and monotonous for a given truncation dimension $\chi$.
 No noticeable singular behavior appears in the entanglement entropy.
 The entanglement entropy is maximal at $R=0$.
 As the $R$ approaches positive or negative infinity, the entanglement entropy approaches the minimum $S=0$, which means that the ground state becomes a product state in the limits.
 Also, for the increment of the truncation dimensions from $\chi=30$ to $\chi=150$,
 the entanglement entropy does not change much for the whole parameter range of $R$.
 Such behaviors of the entanglement entropy implies that
 the ground state of the spin-$2$ system does not undergoes any abrupt substantial change in its nature at a value of the rhombic-type anisotropy $R$ for $D = 4J$.
 To search for a possible nonanalyticity,
 the first- and the second-order derivatives of the entanglement entropy are performed
 and displayed in the insets of Fig. \ref{Fig8}(a).
 Compared with Fig.~\ref{Fig5}(a) for the continuous
 quantum phase transition in the spin-$1$ system,
 the entanglement entropy is maximal at $R=0$ but is not a cusp, i.e.,
 the entanglement entropy according to an increment of the truncation dimension
 does not diverge but converges to a finite value,
 which implies that the spin-$2$ system is not in a critical ground state.
 In the iMPS representation, as shown for the continuous phase transition
 in the spin-$1$ system in Sec. \ref{con-spin1},
 the entanglement entropy diverges as the truncation dimension $\chi$ increases
 at critical points for continuous phase transitions.
 Accordingly, since no explicit abrupt phase separation occurs for the whole parameter space,
 for the spin-$2$ system, the positive biaxial spin nematic state is connected adiabatically
 to the negative biaxial spin nematic state
 by varying the rhombic-type single-ion anisotropy $R$
 from $R/J = -\infty$ to $R/J = \infty$ or vice versa.
 This adiabatic connection between the two orthogonal biaxial
 spin nematic states can be called quantum crossover \cite{Khan,Mao}.

 Figures \ref{Fig8}(b) and \ref{Fig8}(c) show that similar to the entanglement entropy
 in Fig. \ref{Fig8}(a), the WYSIs $I^{x/y/z}$ are continuous and monotonous
 over the entire parameter range.
 In the negative infinite limit of $R/J$ ($R < 0$), the $I^y$ approaches to zero,
 while in the positive infinite limit of $R/J$ ($R > 0$), the $I^x$ approaches to zero.
 In contrast to the $I^{x}$ and $I^{y}$ with $I^x(R)=I^y(-R)$,
 the $I^z$ exhibits a monotonous valley shape, which has the minimum value at $R=0$,
 and thus  has symmetry with respect to $R=0$, i.e., $I^z(R)=I^z(-R)$.
 No noticeable singular behavior is seen in the WYSIs.
 Thus, the first- and the second-order derivatives of WYSIs are performed
 and displayed in the insets of Figs. \ref{Fig8}(b) and \ref{Fig8}(c).
 Similarly, any noticeable significant change for possible phase transitions is
 not observed in the first- and the second-order derivatives of the WYSIs $I^{x/y/z}$.
 To explicitly determine whether quantum phase transitions occur or
 not through the nonanalyticity of WYSIs, their higher-order derivatives should be investigated
 because the nananalyticity of WYSIs has been shown to correspond
 to the critical points of the quantum phase transition well
 for the spin-$1$ system in Sec. \ref{spin-1-system}.
 However, in practice, obtaining reliable higher-order numerical derivatives is
 a very challenging task.
 Accordingly, if an occurrence of quantum phase transition should be judged solely by the nonanalyticity of WYSI,
 in this case it may be invalid to conclude the nonanalyticity of WYSIs
 through such an absence of nonanalyticity of WYSIs only up to the second derivatives of WYSIs.
 However, at least, the behavior of the WYSI is consistent with that of the entanglement entropy by up to the second derivative.
 Therefore, this result shows that the WYSI exhibits a proper behavior
 without any abrupt change for the quantum crossover although the WYSI cannot determine
 whether quantum crossover is or not due to the limitation of numerical accuracy.

%%%%%%%%%%%%%%%%%%%%%%%%%%%%Fig. 9%%%%%%%%%%%%%%%%%%%%%%%%%%%%%%%
\begin{figure}
    \includegraphics[width=0.4\textwidth]{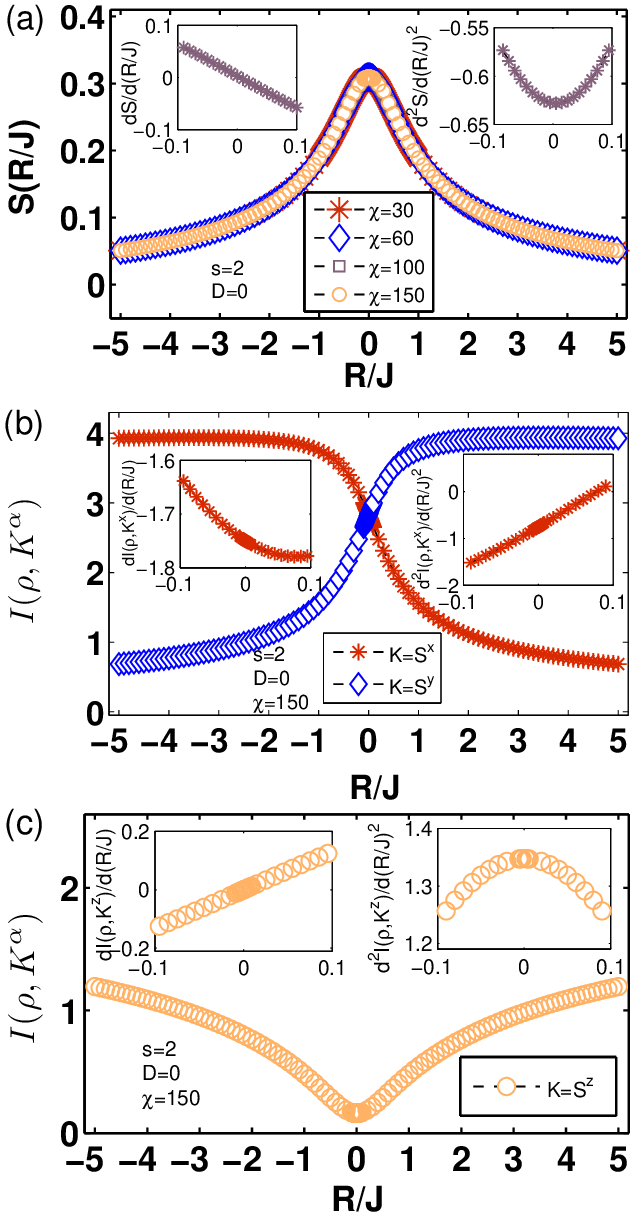}
    \caption{(color online) (a) Entanglement entropy $S$ for various truncation dimensions $\chi=30, 60, 100$ and $150$, (b) WYSI $I^x=I(\varrho, K^x)$ and $I^y=I(\varrho, K^y)$, and (c) $I^z=I(\varrho, K^{z})$
    for $\chi=150$ in the spin-$2$ system with the uniaxial-type anisotropy $D = 0$.
    In the insets, the first- and second-order derivatives of $S$ and $I^{x/y/z}$
    are plotted.
    } \label{Fig9}
     \end{figure}
%%%%%%%%%%%%%%%%%%%%%%%%%%%%%Fig. 9%%%%%%%%%%%%%%%%%%%%%%%%%%%%%%
%
 Let us consider the absence of the uniaxial-type anisotropy $D=0$.
 Figure \ref{Fig9} shows (a) the entanglement entropy $S$,
 (b) the WYSIs $I^{x/y}$ and (c) $I^z$ as a function of $R/J$ for $D = 0$.
 The entanglement entropy and the WYSIs $I^{x/y/z}$ as well as the first-
 and the second-order derivatives behave monotonously without any singular behavior
 very similar to those in Figs. \ref{Fig8}(a)-(c) for $D = 4J$, respectively.
 Overall, the entanglement entropy does not change for the increment of the truncation dimension. This implies that the spin-$2$ system is not in a critical ground state for the whole parameter range.
 Also, the WYSI and their first- and second-derivatives behave very much
 similar to those for $D = 4J$.
 In fact, similar behavior of entanglement entropy and WYSI can be found
 until the anisotropy reaches the critical endpoint at $(R_{CEP},D_{CEP}) =(0, -4.03J)$.
 Consequently, the two biaxial spin nematic states are connected each other
 adiabatically without an explicit abrupt phase transition
 for $D > D_{CEP} = -4.03J$ as the $R/J$ varies from $R/J = -\infty$ to $R/J = \infty$
 or vice versa.

%%%%%%%%%%%%%%%%%%%%%%%%%%%%%%%Fig10%%%%%%%%%%%%%%%%%%%%%%%%%%%%
\begin{figure}
    \includegraphics[width=0.4\textwidth]{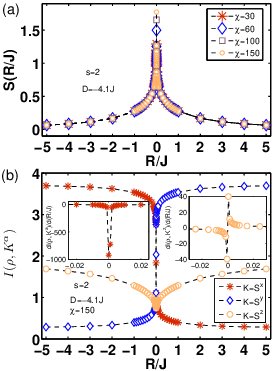}
    \caption{(color online) (a) Entanglement entropy $S$ for various truncation dimensions $\chi=30, 60, 100$ and $150$, and (b) WYSI $I^\alpha=I(\varrho, K^{\alpha})$
    for $\chi=150$ in the spin-$2$ system with the uniaxial-type anisotropy $D = -4.1J$.
    In the insets, the first-order derivatives of $I^{x/y/z}$
    are plotted.
    } \label{Fig10}
     \end{figure}
%%%%%%%%%%%%%%%%%%%%%%%%%%%%%%%%Fig10%%%%%%%%%%%%%%%%%%%%%%%%%%%
%%
 \subsubsection{Continuous quantum phase transition between biaxial spin nematic phases}
 \label{con-spin2}
 Let us consider $D = -4.1J$ in the range of the anisotropy $-4.178J < D < D_{CEP} = -4.03J$.
 In Fig.~\ref{Fig10}, we plot (a) the entanglement entropy $S$ for various truncation dimensions $\chi=30, 60, 100$ and $150$, and (b) the WYSI $I^{x/y/z}$
 for $\chi=150$ as a function of $R/J$ for  $D=-4.1J$.
 Figure \ref{Fig10}(a) shows that
 the entanglement entropy is continuous for a given truncation dimension $\chi$ and its cusp appears at $R=0$.
 Similar to a typical behavior of entanglement entropy in Fig. \ref{Fig5}(a),
 one can notice that the cusp peak value at $R=0$ becomes bigger as the truncation dimension $\chi$ becomes bigger.
 Thus, the spin-$2$ system is in a critical ground state
 because the entanglement entropy diverges in the thermodynamic limit $\chi \rightarrow \infty$, as shown in Fig. \ref{Fig12}(a).
 For the characterization of the critical point,
 the central charge will be calculated in Sec. \ref{phase-rhom2}.
 Compared to the quantum crossover for $D > D_{CEP}$,
 consequently, the cusp behavior of the entanglement entropy at $R=0$ indicates
 a continuous quantum phase transition  occurring between the positive and the negative biaxial spin nematic phases.

%%%%%%%%%%%%%%%%%%%%%%%%%%%%%%Fig11%%%%%%%%%%%%%%%%%%%%%%%%%%%%%
\begin{figure}
    \includegraphics[width=0.4\textwidth]{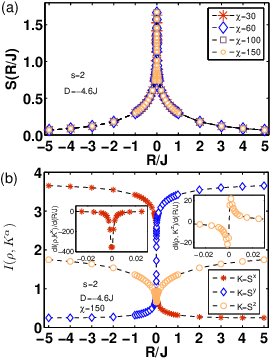}
    \caption{(color online)(a) Entanglement entropy $S$ for various truncation dimensions $\chi=30, 60, 100$ and $150$, and (b) WYSI $I^\alpha=I(\varrho, K^{\alpha})$
    for $\chi=150$ in the spin-$2$ system with the uniaxial-type anisotropy $D = -4.6J$.
    In the insets, the first-order derivatives of $I^{x/y/z}$
    are plotted.
    } \label{Fig11}
     \end{figure}
%%%%%%%%%%%%%%%%%%%%%%%%%%%%%%Fig11%%%%%%%%%%%%%%%%%%%%%%%%%%%%%

 Figure \ref{Fig10}(b) shows that as it should be,
 the WYSIs $I^{x/y/z}$ satisfy the relationships $I^x(R)=I^y(-R)$ and $I^z(R)=I^z(-R)$.
 Compared to those for the quantum crossovers in Figs. \ref{Fig8}(b)-(c) for $D=-4J$ and
 Figs. \ref{Fig9}(b)-(c) for $D=0$, the WYSIs $I^{x/y/z}$ for $D=-4.1J$ seem to undergo a significant change for a very narrow range of $R/J$ around $R=0$.
 An inflection point at $R=0$ is seen in the $I^{x/y}$, while a cusp at the $R=0$
 is observed in the $I^z$.
 The first-order derivatives of the WYSIs $I^x$ and $I^z$ in
 the inset of Fig. \ref{Fig10}(b) exhibit the nonanalyticity of the WYSIs clearer.
 The derivative of the $I^y$ (not presented here)
 exhibits a sharp cusp very similar to those of the $I^x$ in the inset of Fig. \ref{Fig10}(b).
 Actually, the bigger the truncation dimension, the sharper the cusps.
 The derivative of the $I^z$ is discontinuous at the cusp singular point.
 Similar to the characteristic behavior of the WYISs indicating
 the quantum phase transition between $x/y$-ferroquadrupole phases the spin-$1$ system in Fig. \ref{Fig5}(b), such crucial behaviors of the $I^{x/y/z}$ and
 the derivatives indicate an occurrence of continuous quantum phase transition across $R=0$ between the biaxial spin nematic phases in the spin-$2$ system.

%%%%%%%%%%%%%%%%%%%%%%%%%%%%Fig12%%%%%%%%%%%%%%%%%%%%%%%%
\begin{figure}
    \includegraphics[width=0.46\textwidth]{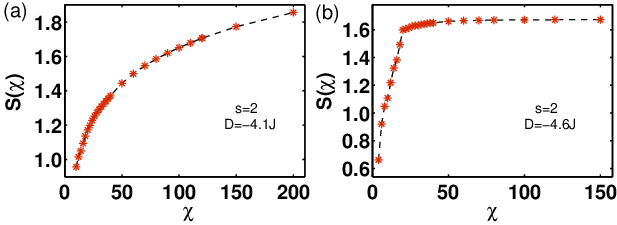}
    \caption{(color online) Entanglement entropy
    as a function of the truncation dimension $\chi$ for (a) $(R,D)=(0,-4.1J)$
    and (b) $(R,D)=(0,-4.6J)$ in the spin-$2$ system.
    } \label{Fig12}
     \end{figure}
%%%%%%%%%%%%%%%%%%%%%%%%%%Fig12%%%%%%%%%%%%%%%%%%%%%%%%

%
 \subsubsection{Discontinuous quantum phase transition between biaxial spin nematic
 phases}
 For $D = -4.6J$, in Fig.~\ref{Fig11}, the entanglement entropy and
 the WYSIs  $I^{x/y/z}$ are plotted as a function of $R/J$.
 Similar to the entanglement entropies in Figs. \ref{Fig5}(a) and \ref{Fig10}(a),
 for a given truncation dimension $\chi$,
 the entanglement entropy is continuous and has the cusp at $R=0$.
 However,
 in contrast to the typical behavior of entanglement entropy
 for the continuous quantum phase transition in Fig. \ref{Fig10}(a),
 as shown in the comparison in Figs. \ref{Fig12}(a) and \ref{Fig12}(b),
 the cusp peak value at $R=0$ in Fig. \ref{Fig11}(b) does not change much and converges as the truncation dimension $\chi$ becomes bigger.
 This fact implies that
 this spin-$2$ system for $D = - 4.6J$ isn't in a critical ground state
 because the entanglement entropy does not diverge
 but converges in the thermodynamic limit $\chi \rightarrow \infty$ in Fig. \ref{Fig12}(b).
 Accordingly, such a cusp, i.e., the nonanalyticity of the entanglement entropy at $R = 0$
 has nothing to do with a critical ground state and a corresponding continuous quantum phase transition.
 Quantum crossover in Sec. \ref{qcrossover}
 has also nothing to do with any cusp of entanglement entropy for a given truncation dimension
 although quantum entanglement converges in the thermodynamic limit $\chi \rightarrow \infty$.
 However, for the discontinuous quantum phase transition in Sec. \ref{dis-spin1},
 it was shown that the entanglement entropy converges in the thermodynamic limit $\chi \rightarrow \infty$ in Fig. \ref{Fig12}(b) and has the discontinues jumps, indicating the nonanalyticity of
 the entanglement entropy, at the critical point for a given truncation dimension.
 Compared to the discontinuity
 of the entanglement entropies at the critical points Figs. \ref{Fig3}(a) and \ref{Fig4}(a),
 the cusp of the entanglement entropy in Fig. \ref{Fig11}(a) is the only difference.
 Both the discontinuity and the cusp of entanglement entropy are nothing but the nonanalyticity of entanglement entropy.
 Hence, the cusp of the entanglement entropy for $D = -4.6J$ in Fig. \ref{Fig11}(a)
 indicates an occurrence of discontinuous quantum phase transition between
 the biaxial spin nematic phases in the spin-$2$ system.

%%%%%%%%%%%%%%%%%%%%%%%Fig13%%%%%%%%%%%%%%%%%%%%%%%%%%%%
\begin{figure}
    \includegraphics[width=0.46\textwidth]{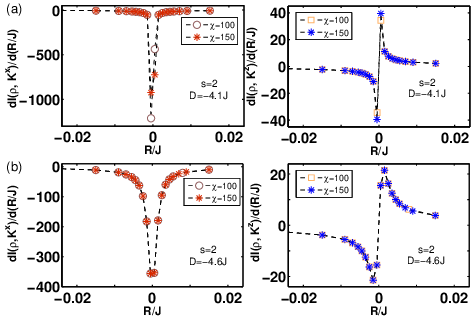}
    \caption{(color online) First-order derivatives of the WYSIs
    as a function of the $R/J$ for (a) $D = -4.1J$
    and (b) $D = -4.6J$ in the spin-$2$ system for $\chi=100$ and $150$.
    } \label{Fig13}
     \end{figure}
%%%%%%%%%%%%%%%%%%Fig13%%%%%%%%%%%%%%%%%%%%%%%%%%%%%%%%

 Let us discuss the WYSIs for this discontinuous quantum phase transition.
 As it should be,
 the WYSIs $I^{x/y/z}$ satisfy the relationships $I^x(R)=I^y(-R)$ and $I^z(R)=I^z(-R)$
 in Fig. \ref{Fig11}(b).
 Similar to the WYSIs $I^{x/y/z}$ for $D=-4.1J$ in Fig. \ref{Fig10}(b),
 Fig. \ref{Fig11}(b) shows an inflection point in the $I^{x/y}$ and a cusp in the $I^z$ at the $R=0$.
 Moreover, similar to the first-order derivatives of the WYSIs $I^{x/y/z}$ for $D=-4.1J$
 in Fig. \ref{Fig10}(b), the first-order derivatives of the WYSIs $I^x$ and $I^z$ in
 the inset of Fig. \ref{Fig11}(b) exhibit the nonanalyticity of the WYSIs.
 To be clear,
 the derivative of the $I^x$ exhibits a cusp very similar to those of the $I^x$ and
 the derivative of the $I^z$ is discontinuous at $R = 0$.
 Actually, the derivative of the $I^y$ (not presented here)
 exhibits a similar cusp to that of the $I^x$.
 The cusp or the discontinuity of the WYSIs depend on the truncation dimension
 for $D = -4.1J$ in Fig. \ref{Fig13}(a),
 while the cusp or the discontinuity of the WYSIs do not depend on the truncation dimension for $D = - 4.6J$ in Fig. \ref{Fig13}(b) although the cusp or the discontinuity of the WYSIs appear. Such nonanalyticities of the WYSIs for  $D = -4.6J$ characterizes the discontinuous quantum phase transition in the spin-$2$ system.

%%
%%%%%%%%%%%%%%%%%%%%%%%%%%%%%%%%%%%%%%%%%%%%%%%%%%%%%%%%%%%
\begin{figure}
    \includegraphics[width=0.4\textwidth]{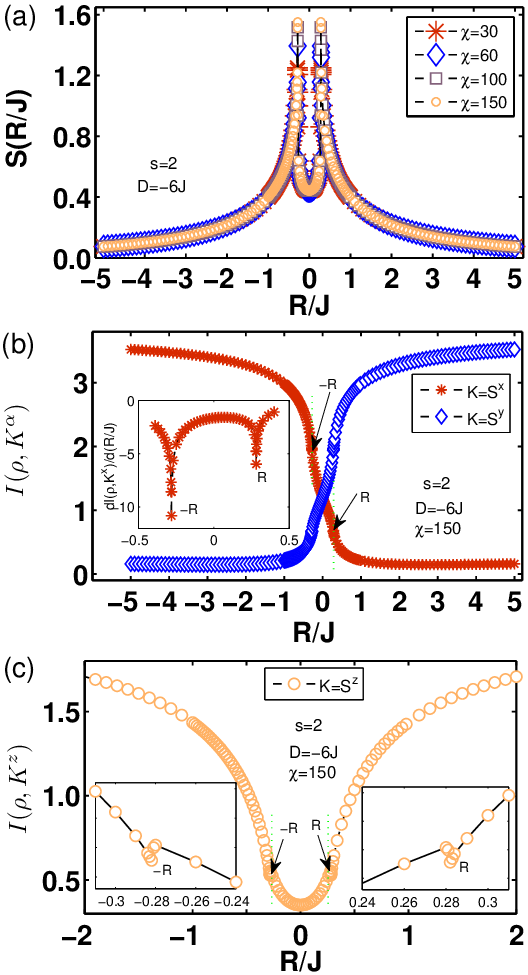}
    \caption{(color online) (a) Entanglement entropy $S$ for various truncation dimensions $\chi=30, 60, 100$ and $150$, (b) WYSI $I^x=I(\varrho, K^x)$ and $I^y=I(\varrho, K^y)$, and (c) $I^z=I(\varrho, K^{z})$
    for $\chi=150$ in the spin-$2$ system with the uniaxial-type anisotropy $D = -6J$.
    In the inset of (b), the first-order derivatives of $I^{x/y}$
    are plotted. In the inset of (c), the plot of $I^z$ is zoomed in near the transition points.
    } \label{Fig14}
     \end{figure}
%%%%%%%%%%%%%%%%%%%%%%%%%%%%%%%%%%%%%%%%%%%%%%%%%%%%%%%%%%%
%%
%
 \subsubsection{Continuous quantum phase transition from biaxial spin nematic phases
  to spin non-nematic phase}
 We have studied the three different types, i.e., quantum crossover, continuous and discontinuous quantum phase transitions between the two biaxial spin nematic phases
 for $D > D_{CEP}$, $ D_M < D < D_{CEP}$, and $D < D_M$, respectively.
 Thus, in this subsection, we consider negatively stronger uniaxial anisotropy, i.e., $ D < D_M$. In fact, for $ D < D_M$,
 the spin-$2$ system undergoes an indirect transition rather than a direct transition from one to the other biaxial spin nematic phases as the rhombic anisotropy varies.
 
 For $D = -6J$, the entanglement entropy $S$ and the WYSIs $I^{x/y/z}$
 are plotted as a function of $R/J$ in Fig.~\ref{Fig14}.
 In contrast to the cases of other parameters $D > D_M$,
 Fig.~\ref{Fig14}(a) shows that the entanglement entropy has a double cusp peak structure.
 As the truncation dimension increases, the two cusps become shaper and higher.
 Moreover, the locations of the two cusps do not change much with the increment of the truncation dimension.
 As was discussed, such a increasing behavior of entanglement entropy with the increment of the truncation dimension indicates occurring a continuous quantum phase transition at the cusps $R = \pm 0.282 J$.
 Accordingly, the occurrence of the two continuous quantum phase transitions 
 implies no direction transition between the two biaxial spin nematic phases and thus another phase, i.e., i.e., antiferromagnetic phase, exist for $-R_c < R < R_c \equiv 0.282 J$.
 The cusps indicate the continuous nonmagnetic to magnetic quantum phase transitions, i.e., between the biaxial spin nematic phase and the antiferromagnetic phase.

%%%%%%%%%%%%%%%%%%%%%%%Fig15%%%%%%%%%%%%%%%%
 \begin{figure}
    \includegraphics[width=0.35\textwidth]{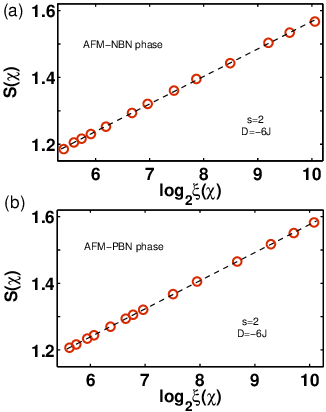}
    \caption{(color online) Entanglement entropy $S(\chi)$ as a function of
    logarithm of correlation length
    $\log_2\xi(\chi)$ for the biquadratic spin-$2$ XY model with the uniaxial-type
    single-ion anisotropy $D$ at the phase boundaries between (a) AFM phase and NBN phase and (b) AFM phase and PBN phase.
    Both the estimate central charges are $c = 1/2$.
    } \label{Fig15}
     \end{figure}
 %%%%%%%%%%%%%%%%%%%%%%Fig15%%%%%%%%%%%%%%%%%

 The two critical points can be characterized by calculating the central charges.
 To do this, we follow the same numerical procedure with the scaling relation in Eq.
 (\ref{entropy1}) in Sec. \ref{phase-rhom}.
 We plot the entanglement entropy $S(\chi)$
 as a function of the logarithmic correlation length $\log_2 \xi(\chi)$
 for the continuous quantum phase transitions from the antiferromagnetic phase to
 (a) the negative biaxial spin nematic phase
  at $R=-R_C$ or (b) the positive biaxial nematic phase $R=+R_c$ in Fig. \ref{Fig15}.
 By employing the numerical fitting function $S(\chi) = \frac{c}{6}  \log_2 \xi(\chi) + S_0$,
 we obtain the numerical fitting coefficients
 (a) $c/6=0.082(1)$ and $S_0=0.743(8)$ at $R=-R_c$ and
 (b) $c/6=0.0844(4)$ and $S_0=0.732(3)$ at $R=+R_c$.
 Accordingly,
 the central charges are estimated as $c=0.495(6)$ at $R=-0.282J$ and $c=0.506(2)$ at $R=+0.282J$, respectively.
 Both the phase boundaries between the antiferromagnetic phase and the negative or positive
 biaxial spin nematic phases have the central charge $c \simeq 1/2$, which belongs to the Ising universality class.

 Figures \ref{Fig14}(b) and \ref{Fig14}(c) show
 that similar to the cases of other chosen parameters,
 the WYSIs satisfy the relationships  $I^x(R)=I^y(-R)$ and $I^z(R)=I^z(-R)$.
 However, the $I^{x/y}$ has two inflection points and $I^z$ has two cusps
 at $R = \pm R_c$.
 The two inflection points can be confirmed clearly
 as the corresponding two cusps in the firs-order derivatives
 of the $I^x$ in the inset of Fig. \ref{Fig14}(b).
 The two cusps can be observed clearly in the closeup around $R = \pm R_c$
 in the inset of Fig. \ref{Fig14}(c).
 Accordingly, the WYSIs $I^{x/y/z}$ 
 capture the two continuous nonmagnetic to magnetic quantum phase transitions.
 Similar to the entanglement entropy,
 the WYSIs reveal the existence of another distinct phase between the
 positive (negative) biaxial spin nematic phases for $- R_c < R < R_c$.

 As we discussed, for $D > D_{CEP}$, no explicit phase transition occurs. Explicit quantum phase transition occurs for $D_M < D < D_{CEP}$ and thus one critical point at $R=0$
 exits for a given
 uniaxial-type single-ion anisotropy $D$.
 As the anisotropy decreases than the multicritical point $D = D_M$, the critical point split into two critical points $R=\pm Rc$.
 The locations of two critical points $R_c$
 depend on the strength of the uniaxial-type single-ion anisotropy $D$.
 The stronger the negative anisotropy, the larger the $R_c$.
 Such a dependence between the anisotropies $R$ and $D$ for the two phase boundaries
 seems to be linear,
 as shown in the phase diagram of the spin-$2$ system in Fig. \ref{Fig2}.
 The two linear phase boundaries given by $R=\pm R_c$ clearly separate the two biaxial spin
 nematic phases from the antiferromagnetic phase that appears in between.
 The phase boundaries can be numerically fitted with the linear fitting function $D_c/J = \pm\, a\, R_c/J + b$. The performed fitting gives the fitting coefficients $a = -3.6(1)$ and $b = -4.95(9)$. The estimate value of the multicritical point from the fitting
 is given as  $D_M =-4.95(9)J$. The estimate value from the entanglement entropy $S$
 and the WYSIs $I^{x/y/z}$ is $ D_M = -4.94J$. The discrepancy between the two estimates
 are from the nonlinearity of the actual phase boundary near the multicritical point.

%%
%%%%%%%%%%%%%%%%%%%%%%%%%%%%Fig16%%%%%%%%%%%%%%%%%%%%%%%%%%%%%%%
\begin{figure}
    \includegraphics[width=0.4\textwidth]{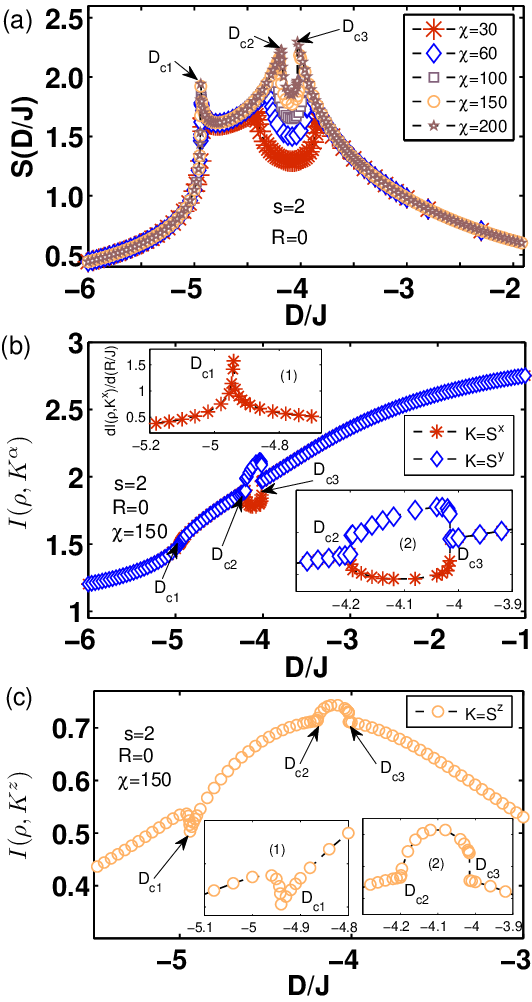}
    \caption{(color online) (a) Entanglement entropy $S$ for various truncation dimensions $\chi=30, 60, 100$, $150$ and $200$, (b) WYSI $I^x=I(\varrho, K^x)$ and $I^y=I(\varrho, K^y)$, and (c) $I^z=I(\varrho, K^{z})$
    as function of $D/J$
    for $\chi=150$ in the spin-$2$ system with the rhombic-type anisotropy $R = 0$.
   In the top-right inset of (b), the first-order derivatives of $I^{x/y}$
    are plotted. In the bottom-left inset of (b), the plots of $I^{x/y}$ are zoomed in near
    the transition points $D_{c2}$ and $D_{c3}$.  In the insets of (c), the plot of $I^z$ is zoomed in near the transition points $D_{c1}$, $D_{c2}$ and $D_{c3}$.
    } \label{Fig16}
     \end{figure}
%%%%%%%%%%%%%%%%%%%%%%%%%%%%Fig16%%%%%%%%%%%%%%%%%%%%%%%%%%%%%%%

 \subsubsection{Phase diagram of the biquadratic spin-$2$ XY chain with the uniaxial-type single-ion anisotropy}
 \label{phase-rhom2}
 As we have discussed in Sec. \ref{phase-rhom}, for the spin-$1$ system without the rhombic-type single-ion anisotropy
 $R=0$, there are the two distinct phases, i.e., the critical phase
 and the $z$-ferroquadruple phase.
 In the absence of the rhombic-type single-ion anisotropy $R=0$,
 the biquadratic spin-$2$ XY model Hamiltonian with the uniaxial single-ion anisotropy
 has the same form of Eq. (\ref{ham2}) by replacing the spin-$1$ operator with the spin-$2$
 operator.
 In the limit of $D/J \rightarrow \infty$ with $J > 0$, the ground state of the spin-$2$ system is the product state of $|S^z_i =0\rangle$. This phase is referred to as the large-$D$ phase.
 On the contrary, in the limit of $D/J \rightarrow -\infty$ with $J > 0$, the ground state
 of the spin-$2$ system is in an Ising-like antiferromagnetic phase characterized by finite
 staggered magnetizations. The antiferromagnetic phase has a doubly degenerate ground state.

%%%%%%%%%%%%%%%%%%%%%%%%Fig17%%%%%%%%%%%%%%%
 \begin{figure}
    \includegraphics[width=0.4\textwidth]{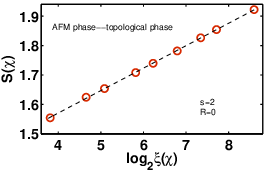}
    \caption{(color online) Entanglement entropy $S(\chi)$ as a function of
    logarithm of correlation length,
    $\log_2\xi(\chi)$, for the biquadratic spin-$2$ XY model with the uniaxial-type
    single-ion anisotropy $D$ at the transition point $D_{c1}=D_M$  between the antiferromagnetic phase and
   the topological phase for the truncation dimension from $\chi=20$ to $\chi=150$.
    The estimate central charge is $c =0.458(6)$.
    } \label{Fig17}
     \end{figure}
 %%%%%%%%%%%%%%%%%%%%%%Fig17%%%%%%%%%%%%%%%%%

 In Fig. \ref{Fig16},
 we plot the entanglement entropy $S$ and the WYSIs $I^{x/y/z}$ as a function of $D/J$ for $R=0$.
 The entanglement entropy shown in Fig. \ref{Fig16}(a) has a three cusp peak structure.
 As the truncation dimension increases, all the peaks become shaper and higher.
 Also, the left cusp $D_{c1}$ does not change its location significantly
 but the middle $D_{c2}$ and right $D_{c3}$ cusps move towards each other.
 The $D_{c1}$ and $D_{c3}$ correspond to the multicritical point $D_M$ and the critical end-point $D_{CEP}$, respectively.
 As the truncation dimension increases,
 the entanglement entropy of $D_{c2} < D < D_{c3}$ increases, while the entanglement entropy of the parameter range except $D_{c2} < D < D_{c3}$ remains unchanged. This implies
 that the ground state of the spin-$2$ system is critical for $ D_{c2} < D < D_{c3}$.
 Consequently, with the known phases, i.e., the antiferromagnetic phase for $D < D_{c1}$ and the large-D phase for $D > D_{c3}$, a critical phase and another gapful phase appear
 for $D_{c2} < D <D_{c3}$ and $D_{c1} < D <D_{c2}$, respectively.

%%%%%%%%%%%%%%%%%%%%%Fig18%%%%%%%%%%%%%%%%%%%%
    \begin{figure}[b]
    \includegraphics[width=0.4\textwidth]{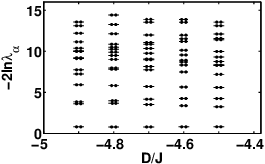}
    \caption{(color online) Entanglement spectrum of Hamiltonian in Eq. (\ref{ham2})
    for the biquadratic spin-$2$ XY model with the uniaxial-type single-ion anisotropy
    for various value of $D/J$ in the topological phase with $R/J=0$.
    The dots show the first $30$ coefficients of the entanglement spectrum $-2 \ln \lambda_\alpha$ for $\chi=30$.
    } \label{Fig18}
     \end{figure}
%%%%%%%%%%%%%%%%%%%%%Fig18%%%%%%%%%%%%%%%%%%%%%%%%%

 Similar to the nonanalyticity of the entanglement entropy,
 the WYSIs in Figs. \ref{Fig16}(b) and \ref{Fig16}(c) respectively show that the $I^{x/y}$ has one inflection and two kinks, while the $I^z$ has one cusp and two kinks.
 The cusp first-order derivative of the $I^x$ manifest the nonanalyticity
 of the $I^x$ at $D_{c1}$ in the inset of Fig. \ref{Fig16}(b).
 The closeups of the $I^{x/y}$ and the $I^z$ make the kinks $D_{c2}$ and $D_{c3}$ clear
 in the insets of Figs. \ref{Fig16}(b) and \ref{Fig16}(c), respectively.
 The three singular points indicate occurring continuous quantum phase transitions
 with the transition points, e.g., $D_{c1} = -4.94J$, $D_{c2} = -4.178J$ and
 $D_{c3}=-4.03J$ for the truncation dimension $\chi=200$.
 The WYSIs also reveal the existence of another gapful phase appear for $D_{c1} < D <D_{c2}$.

%%%%%%%%%%%%%%%%%%%%%%%%%%Fig19%%%%%%%%%%%%%%%%%%%%%%%%%%%%%
   \begin{figure}
    \includegraphics[width=0.4\textwidth]{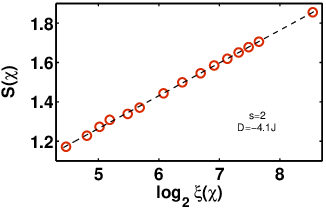}
    \caption{(color online)  Entanglement entropy $S(\chi)$ as a function of
    logarithm of correlation length,
    $\log_2\xi(\chi)$, for the biquadratic spin-$2$ XY model with the uniaxial-type
    single-ion anisotropy $D$ for $D = -4.1J$ with the truncation dimension $\chi$ from $20$ to $200$.
    The estimate central charge is $c \simeq 1$  in the critical region $D_{c2} <D <
    D_{c3}$.
    } \label{Fig19}
     \end{figure}
%%%%%%%%%%%%%%%%%%%%%%%%FIg19%%%%%%%%%%%%%%%%%%%%%%%%%%%%%%%

%
 Thus, let us first discuss about the continuous quantum phase transition from the antiferromagnetic phase to the gapful phase across the multicritical point $D_M = D_{c1}$.
 In order to calculate the central charge, we plot the entanglement entropy as a function of logarithm  of correlation length $\log_2 \xi$ in Fig. \ref{Fig17}.
 The numerical fitting is performed to estimate the central charge $c$ with the fitting function $S(\chi)= \frac{c}{6} \log_2 \xi(\chi)+S_0$.
 The numerical fittings give  $c/6=0.076(1)$ and $S_0=1.265(6)$.
 Compared to the transitions from the positive/negative biaxial spin nematic phases
 to the antiferromagnetic phase belong to the Ising universality class with $c \simeq 1/2$,
 the transition from the gap phase to the antiferromagnetic phase
 has the estimate central charge $c=0.458(6)$ smaller than that of the Ising universality class.

 Nothing is known yet except that the ground state of the spin-$2$ system for $D_{c1} < D < D_{c2}$  is not in the critical state and gapful,
 compared relatively to the antiferromagnetic, critical phase and large-$D$ phases.
 In order to understand more about the spin-$2$ system for $D_{c1} < D < D_{c2}$,
 let us thus consider the entanglement spectrum \cite{Li}, i.e. the set of the eigenvalues $\{\lambda^2_\alpha \}$ of the reduced density matrix $\varrho_L$ or $\varrho_R$.
Compared to the entanglement entropy $S(\{\lambda^2_\alpha\})$ being a single number in Eq. (\ref{SS}), the entanglement spectrum can contain information, not included
 in the entanglement entropy, such as a symmetry protected phase
 \cite{Chen,Pollmann1,Zhang}.
 In Fig.~\ref{Fig18}, we plot the entanglement spectrum $-2\ln \lambda_\alpha$ with
 the first $30$ levels for various parameters $D=-4.5J, -4.6J, -4.7J, -4.8J$ and $-4.9J$
 in the biquadratic spin-$2$ XY model with the uniaxial-type single-ion anisotropy.
 For the parameter range except $D_{c1} < D  < D_{c2}$, nondegenerate entanglement spectrum appears (not presented here).
 For $D_{c1} < D < D_{c2}$, on the other hand, the entire entanglement spectrum is
 doubly degenerate. Such a degenerate structure in entanglement spectrum is a characteristic property of topological phase.
 We will discuss an order parameter characterizing this topological phase in Sec. \ref{top}.

 For $ D_{c2} < D < D_{c3}$, the critical ground state can be characterized by calculating
 a central charge . Thus, in Fig.~\ref{Fig19},
 we plot the entanglement entropy $S(\chi)$ as the correlation
 length $\log_2 \xi(\chi)$ for  $D=-4.1J$.
 The fitting coefficients are given by $c/6=0.167(3)$ and $S_0=0.43(2)$.
 The resulting central charge is $c=1.00(2)$ in the $D_{c2} < D < D_{c3}$ region.

%%%%%%%%%%%%%%%%%%%%%%%%%%%%%%%Fig20%%%%%%%%%%%%%%%%%%%%%%%%%%%%%%%
\begin{figure}
    \includegraphics[width=0.4\textwidth]{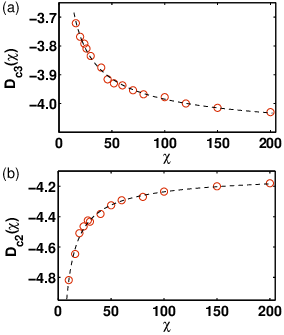}
    \caption{(color online) Transition points (a) $D_{c3}(\chi)$ and (b) $D_{c2}(\chi)$ as a function of the truncation dimension $\chi$. An extrapolations are performed to estimate the critical points in the thermodynamics.
    } \label{Fig20}
     \end{figure}
%%%%%%%%%%%%%%%%%%%%%%%%%%%%%%%%Fig20%%%%%%%%%%%%%%%%%%%%%%%%%%%%%%%%%%%

 As we discussed, the larger the truncation dimension, the closer the two kinks $D_{c2}$ and $D_{c3}$ are. We thus plot the two kink points $D_{c2}(\chi)$ and $D_{c3}(\chi)$
 as a function of the truncation dimension $\chi$ in Fig. \ref{Fig20}.
 Fig. \ref{Fig20}(a) shows the kink $D_{c3}$ decreases as the truncation dimension $\chi$ increases, while Fig. \ref{Fig20}(b) shows the kink $D_{c2}$ increases as the truncation dimension $\chi$ increases. To determine how much the two kinks $D_{c2}$ and $D_{c3}$ get closer in the thermodynamic limit, the numerical extrapolations are performed on the two kinks with the fitting function $D_c(\chi)/J=a\chi^b+D_c(\infty)/J$ \cite{Tagliacozzo}.
 From the extrapolations, the critical points are obtained as $D_{c3}(\infty)=-4.11(6)J$
 with the fitting coefficients $a_{c3}=-4(1)$ and $b_{c3}=-0.7(1)$ in Fig. \ref{Fig20}(a), and $D_{c2}(\infty)=-4.13(5) J$ with the fitting coefficients $a_{c2}=2.0(6)$ and $b_{c2}=-0.6(1)$ in Fig. \ref{Fig20}(b). The estimates give the critical points as $D_{c2}(\infty)=-4.13(5) J$ and $D_{c3}(\infty)=-4.11(6)J$ and thus the critical phase may occur for $-4.13(5)J < D < -4.11(6)J$. However, with $\Delta D_{c2} = 0.05J$ and $\Delta D_{c3} = 0.06J$, the two fitting errors overlap quiet a bit. The error overlap may suggest that there is no critical region between the two critical points and the two critical points merge into one critical point. At this stage, it is beyond the accuracy of our numerical calculations to clearly determine whether the critical region exists or becomes just a phase transition point. Regardless of whether we can make the conclusion about the existence of critical phase or not,in sharp contrast to the spin-$1$ system with two distinct phases in Fig. \ref{Fig6}, the biquadratic spin-$2$ XY model with the uniaxial-type single-ion anisotropy $D$ has a very different phase diagram including a topological phase and the antiferromagnetic phase in Fig. \ref{Fig16}, although the spin-$1$ and -$2$ systems have the same form of the Hamiltonian (\ref{ham2}).

%%%%%%%%%%%%%%%%%%%%%%%%%%%%%%%%%%%%%%%%%%%%%%%%%%%%%%%%%%
\begin{figure}
    \includegraphics[width=0.4\textwidth]{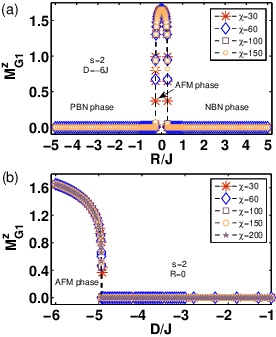}
       \caption{(color online) Staggered magnetization $M^z_{G_1}=\langle S^z_j-S^z_{j+1} \rangle_{G_1}/2$
       as a function of (a) $R/J$ for $D= -6J$ or (b) $D/J$ for $R= 0$ with various truncation dimensions $\chi$.
       Here, to calculate the staggered magnetization, we have chosen one of the two ground states $|\psi_{G_1}\rangle$ and $|\psi_{G_2}\rangle$ in the antiferromagnetic phase.
       The other staggered magnetization
       from the ground state $|\psi_{G2}\rangle$ satisfies $ M^z_{G_1} = - M^z_{G_2}$.
    } \label{Fig21}
     \end{figure}
%%%%%%%%%%%%%%%%%%%%%%%%%%%%%%%%%%%%%%%%%%%%%%%%%%%%

%%%%%%%%%%%%%%%%%%%%%%%%%%%%%%%%%%
\section{Quadrupole orders and quantum spin nematic phase transitions}
 \label{section4}
 The spin-exchange coupling between quantum spins does not always cause magnetic phases of materials with long-range magnetic order. Such nonmagnetic materials have short-range spin correlations. The ground state of the spin-$1$ system does not have local magnetization,
 i.e., $\langle S^\alpha_j \rangle =0$, for all the parameter ranges of the anisotropies $R$ and $D$, as shown in the phase diagram
 in Fig. \ref{Fig1}, while the spin-$2$ system has the antiferromagnetic phase for relatively much bigger $D$ than $R$ in Fig. \ref{Fig2}.
 The antiferromagnetic phase has $\langle S^{x/y}_j \rangle =0$
 and $\langle S^{z}_j \rangle \neq 0$.
 The oriented and ordered spin state is induced by spontaneous spin-rotational symmetry breaking. The antiferromagnetic phase is characterized by the magnetic order parameter, i.e., the staggered magnetization $\langle S^z_j-S^z_{j+1} \rangle/2$.
 In fact, the ground state is doubly degenerate due to the symmetry breaking in the antiferromagnetic phase. The iMPS approach enables to obtain the two degenerate ground states $|\psi_{G_1}\rangle$ and $|\psi_{G_2}\rangle$ in the antiferromagnetic phase by using the quantum fidelity \cite{Su13}. With the first ground state $|\psi_{G_1}\rangle$,
 we plot the staggered magnetization, i.e., $ M^z_{G_1}=\left( \langle S_j^z\rangle_{G_1} -\langle S_{j+1}^z\rangle_{G_1}\right)/2$
 as a function of (a) $R/J$ for $D = -6J$ or (b) $D/J$ for $R = 0$ in Fig. \ref{Fig21}.
 As it should be, $ M^z_{G_1}=-M^z_{G_2}$ (not presented here).
 For $D = -6J$ and $\chi=150$, Fig. \ref{Fig21}(a) shows that the staggered magnetization
 is finite for $ -0.282 J < R < 0.282J$ and is zero for $R > 0.282J$ and $R < -0.282J$.
 The estimate points $R = \pm 0.282 J$ from the staggered magnetization correspond to the cusps of the entanglement entropy in Fig. \ref{Fig14}(a),
 the two inflection points of $I^{x/y}$ in Fig. \ref{Fig14}(b), and the two cusps of $I^z$
 in Fig. \ref{Fig14}(c), respectively. For $R=0$ and $\chi=150$ in Fig. \ref{Fig21}(b), the staggered magnetization is finite for $D < -4.94J$, otherwise it is zero. The critical points $D = -4.94J$ from the order parameter correspond to the cusp of the entanglement entropy in Fig. \ref{Fig16}(a), the inflection point of $I^{x/y}$ in Fig. \ref{Fig16}(b), and the cusp of $I^z$ in Fig. \ref{Fig16}(c), respectively. It is found that for other truncation dimensions, similar correspondences between the transition points detected by the three different physical quantities. Consequently, the nonanalytical points of the entanglement entropy and the WYSIs $I^{x/y/z}$ are consistent with the transition points measured by the order parameter of the antiferromagnetic phase.

 Except for the $(R/J,D/J)$-parameter space corresponding to the antiferromagnetic phase of the spin-$2$ system, shown in Fig. \ref{Fig2}, all the ground states of our spin-$1$ and -$2$ systems have zero magnetization, i.e., $\langle S^\alpha_j\rangle=0$,
 and thus spin nematic ground states with multiple-spin ordering can appear. Such a spin nematic state preserves time-reversal symmetry but breaks spin-rotational symmetry due to the average value of multiple spin. The characterization of spin nematic phases for our quantum biquadratic XY model with rhombic-type single-ion anisotropy $R$ and uniaxial-type single-ion anisotropy $D$ can be performed by calculating and analyzing spin quadrupole moments from the quadrupole tensor operator \cite{Mao, Lai} defined as
\begin{equation}
 Q^{\alpha\beta}_j=\frac{1}{2}\left(S^{\alpha}_jS^{\beta}_j+S^{\beta}_jS^{\alpha}_j\right) - \frac{1}{3} \mathbf{S}^2_j \delta^{\alpha\beta}
 \label{Quadrupolar}
\end{equation}
 with the Kronecker delta $\delta^{\alpha\beta}$ and $\alpha,\beta \in  \{x, y, z\}$ at site $j$. The spin quadrupole tensor operator is symmetric and traceless rank-$2$ tensor, i.e., $Q^{\alpha\beta}_j = Q^{\beta\alpha}_j$ and $\sum_{\alpha} Q^{\alpha\alpha} = 0$, which leads to only five independent components. In fact, the ground states of the spin-$1$ and -$2$ systems have $\langle Q^{\alpha\beta}_j\rangle =0$ for $\alpha \neq \beta$.
 Due to the tracelessness of the tensor, also the quadrupole tensor operator has
 the two actual independent components, i.e., $Q_j^{x^2-y^2}=Q_j^{xx}-Q_j^{yy}$ and $Q^{3z^2-r^2}_j=3Q_j^{zz}$, which can identify quantum spin nematic phases in our spin chains.

\subsection{Uniaxial spin nematic phases (Ferroquadrupole phases) for spin-$1$ system}
\label{Ferro1}
 Let us consider a case of finite value of the uniaxial-type single-ion anisotropy $D$.
 In the limit of very large negative (positive) $R/J \gg D/J$, the rhombic-type single-ion
 anisotropy $R$ becomes predominant in the Hamiltonian (\ref{ham1})
 and thus the local spin state
 is to be the lowest-energy state of the anisotropy term
 $R[(S^{x}_j)^2-(S^{y}_j)^2]$ ($R[(S^{y}_j)^2-(S^{x}_j)^2]$),
 i.e.,
 $|S^y_j=0 \rangle$ ($|S^x_j=0 \rangle$).
 The lowest-energy state implies that the local spin fluctuates in the $zx$ ($yz$) plane \cite{Tzeng}.
 Accordingly, the ground state becomes the product state of $|S^y_j=0 \rangle$ ($|S^x_j=0 \rangle$), i.e.,
 $|\psi_G(R/J \rightarrow -\infty)\rangle = \prod_j \otimes |S^y_j=0\rangle$ ($|\psi_G(R/J \rightarrow \infty)\rangle= \prod_j \otimes |S^x_j=0 \rangle$).
 Thus, the quadrupole moments are given as
 $\langle Q^{xx}_j\rangle = \langle Q^{zz}_j \rangle = -\frac{1}{2}\langle Q^{yy}_j \rangle$
 ($\langle Q^{yy}_j\rangle = \langle Q^{zz}_j \rangle = -\frac{1}{2}\langle Q^{xx}_j \rangle$)
 with $\langle Q^{yy}_j \rangle = -2/3$ ($\langle Q^{xx}_j \rangle = -2/3$), which implies that the ground state is a uniaxial spin nematic state.
 For the limit of very large negative (positive) $R/J \gg D/J$, the spin-$1$ system
 is in the $y$($x$)-ferroquadrupole phase $\langle Q^{x^2-y^2}_j\rangle > 0$ ($\langle Q^{x^2-y^2}_j\rangle < 0$) \cite{Mao}.
 We will discuss how the transition between the spin nematic phases occurs
 as the $R/J$ varies from the negative large value to the positive large value
 for a given uniaxial-type single-ion anisotropy $D$ in the following Secs. \ref{S1QDis}
 and \ref{S1QCon}.

\subsubsection{Discontinuous spin nematic to nematic phase transitions and quadrupole moments}
 \label{S1QDis}
\begin{figure}
    \includegraphics[width=0.4\textwidth]{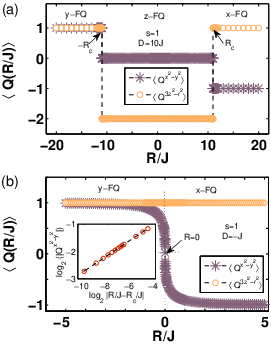}
    \caption{(color online) Quadrupole order parameters $\langle Q_j^{x^2-y^2} \rangle$ and
    $\langle Q_j^{3z^2-r^2}\rangle$ for (a) $D=10J$ and (b)$D=-J$ in the spin-$1$ system with $\chi=150$. In the inset of (b), the least-squares fit of the data near the critical point     $R_c=0$ with $\chi=150$. The estimate critical exponent is $\beta=0.293(3)$.}
     \label{Fig22}
     \end{figure}

 Let us consider the case of $D=10J$.
 In Fig.~\ref{Fig22}(a), we plot the quadrupole moments $\langle Q^{x^2-y^2}_j
 \rangle$ and $\langle Q^{3z^2-r^2}_j\rangle$ as a function of $R/J$ with $\chi=150$.
 As the $R/J$ increases from the negative large value to the positive large value,
 the $\langle Q^{x^2-y^2}_j\rangle$ decreases and undergoes a sequential sharp down at $R=-10.972J$ and $R=+10.972J$.
 While the $\langle Q^{3z^2-r^2}_j\rangle$ undergoes sequentially a sharp down at $R = -10.972J$
 and a sharp back up at $R = +10.972J$.
 The discontinuous points of the quadrupole moments are consistent with the phase transition points detected by the entanglement entropy in Fig. \ref{Fig3}(a) and the WYSIs in Fig. \ref{Fig3}(b).
 These two transition points indicate that there is another phase between
 the $y$-ferroqudarupole phase and the $x$-ferroquadrupole phase for $-10.972J < R < +10.972J$.
 The phase for for $-10.972J < R < +10.972J$ can be identified by
 $\langle Q^{x^2-y^2}_j\rangle =0$ and $\langle Q^{3z^2-r^2}_j\rangle = -2$ as shown in  Fig. \ref{Fig22}(a). The results lead to $\langle Q^{xx}_j\rangle=\langle Q^{yy}_j\rangle=-\langle Q^{zz}_j \rangle/2$, which implies that the local spin fluctuates only in the $xy$ plane.
 The ground state is a uniaxial spin nematic state~\cite{Papanicolaou1988,Chubukov90} and the spin-$1$ system is in the $z$-ferroquadrupole phase. As a result, the three distinct spin nematic phases can be identified as the $x$-ferroquadrupole phase for $R > 10.972J$, the $z$-ferroquadrupole phase for $ -10.972J < R < 10.972J$, and the $y$-ferroquadrupole phase for $R < -10.972J$.

%%%%%%%%%%%%%%%%%%%%%%%Fig22%%%%%%%%%%%%%%%%%%%%%%%%%%
%%%%%%%%%%%%%%%%%%%%%Fig22%%%%%%%%%%%%%%%
 %
 As was discussed in Sec. \ref{dis-spin1}, similar to the entanglement entropy
 and the WYSIs, the locations of the two discontinuous points of the quadrupole moments are dependent on the strength of the uniaxial-type single-ion anisotropy $D$ bigger than $D = -0.612J$ at the multicritical point. By using the discontinuous points of the quadrupole moments depending on the $D/J$, one can estimate the same phase boundaries with that from the entanglement entropy in Sec. \ref{dis-spin1}. For $D > D_M$, the spin-$1$ system has the three distinct spin nematic phases and the discontinuous spin nematic phase transitions occur from the $z$-ferroquadrupole phase to the $x$-ferroquadrupole phase or to the $y$-ferroquadrupole phase, as shown in the spin-$1$ phase diagram in Fig. \ref{Fig1}.

\subsubsection{Continuous spin nematic to nematic phase transitions and quadrupole moments}
\label{S1QCon}
 Let us consider the uniaxial-type single-ion anisotropy $D$ smaller than $D_M$, for
 instance, $D = -J$. The quadupole moments $\langle Q^{x^2-y^2}_j \rangle$ and $\langle Q^{3z^2-r^2}_j\rangle$ are plotted as a function of $R/J$ with the truncation dimension $\chi=150$ in Fig.~\ref{Fig22}(b). As the $R/J$ increases from the negative large value to the positive large value, the $\langle Q^{x^2-y^2}_j\rangle$ starts to decrease from $1$ roughly around $R \sim -2J$, decreases relatively rapidly around the inflection point at $R=0$, and saturates to $-1$ around $R \sim 2J$. In contrast, the $\langle Q^{3z^2-r^2}_j\rangle$ does not change with the strength of the uniaxial-type
 single-ion anisotropy, i.e., $\langle Q^{3z^2-r^2}_j\rangle = 1$ for the whole parameter range of $R/J$. The spin-$1$ system is in the $x$($y$)-ferroquadrupole phase for $ R > 0$ ($ R< 0$), i.e., $\langle Q^{x^2-y^2}_j\rangle > 0$ ($\langle Q^{x^2-y^2}_j\rangle < 0$).

 As for the order parameters of the $x$- and the $y$-ferroquadrupole phases,
 a log-log plot of the $\langle Q^{x^2-y^2}_j\rangle$ near the transition point $R=0$ is displayed in the inset of Fig.~\ref{Fig22}(b). The inset shows a scaling behavior of the quadrupole moment the $\langle Q^{x^2-y^2}_j\rangle$ near the transition point $R=0$.
 Such a behavior of quadrupole moment can be described by the power law, i.e.,
 $\langle Q^{x^2-y^2}_j\rangle \propto |R/J-R_c/J|^\beta$, where $\beta$ is the characteristic
 critical exponent. To estimate the critical exponent, the least-squares fit of the data near the critical point $R=0$ and thus the estimate critical exponent is given as $\beta=0.293(3)$ for the quadrupole moment. Accordingly, the quadrupole order parameter characterizes the continuous spin nematic to nematic phase transition between the $x$- and the $y$-ferroquadrupole phases. This result is consistent with that of the entanglement entropy and the WYSIs.

%%%%%%%%%%%%%%%%%%%%%%%%%%%%%Fig. 23%%%%%%%%%%%%%%%%%%%%
\begin{figure}
    \includegraphics[width=0.4\textwidth]{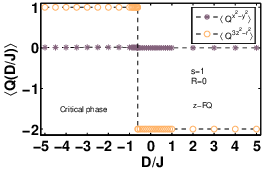}
    \caption{(color online) Quadrupole order parameters $\langle Q_j^{x^2-y^2} \rangle$ and
    $\langle Q_j^{3z^2-r^2}\rangle$ at $R=0$ in the spin-$1$ system with $\chi=150$.
     }
     \label{Fig23}
     \end{figure}
%%%%%%%%%%%%%%%%%%%%%%%%%%%%Fig. 23%%%%%%%%%%%%%%%%%%%%
%%
 \subsubsection{Quadrupole moments in the biquadratic spin-$1$ XY chain with the uniaxial-type single-ion anisotropy}
 Let us consider the spin-$1$ chain without the rhombic-type single-ion anisotropy $R = 0$.
 When the strength of the uniaxial-type single-ion anisotropy $D$ varies,
 we plot the quadrupole moments $\langle Q^{x^2-y^2}_j\rangle$ and $\langle Q^{3z^2-r^2}_j\rangle$ as a function of $D/J$ in Fig.~\ref{Fig23}.
 Noticeably, $\langle Q^{x^2-y^2}_j \rangle =0$, i.e., $\langle Q^{xx}_j\rangle=\langle Q^{yy}_j\rangle$ do not depend on the uniaxial-type single-ion anisotropy $D/J$ in Fig. \ref{Fig23}. In contrast, Fig. \ref{Fig23} shows that
 $\langle Q^{3z^2-r^2}_j\rangle = 1$ for $D < -0.612J$
 and $\langle Q^{3z^2-r^2}_j\rangle = -2$ for $D > -0.612J$.
 The results mean that for the whole parameter range $D/J$,
 $\langle Q^{xx}_j\rangle=\langle Q^{yy}_j\rangle=-\langle Q^{zz}_j \rangle/2$
 and the ground state is a uniaxial spin nematic state.
 For $D > -0.612J$, the spin fluctuation are given as  $\langle (S^{x/y}_j)^2\rangle=1$ and $\langle (S^{z}_j)^2\rangle=0$, which implies that the local spin flutuates like a disk in the $xy$ plane and have zero amplitude along the $z$ axis, i.e., the $z$-ferroquadrupole state. Whereas for $D < -0.612J$, the spin fluctuations are given as
 $\langle (S^{x/y}_j)^2\rangle=1/2$ and $\langle (S^{z}_j)^2\rangle=1$.
 Compared to the $z$-ferroquadrupole state, this state has stronger amplitude of
 spin fluctuation along the $z$ axis with a relatively weaker isotropic fluctuation in the $xy$ plane. Such a drastic change of spin fluctuation at $D = -0.612J$ originates the discontinuous spin nematic phase transition, as shown in Fig. \ref{Fig23}.

%%%%%%%%%%%%%%%%%%%%%%%%%%Fig24%%%%%%%%%%%%%%%%%%%%%%%%%%%%%%%%%%
\begin{figure}
    \includegraphics[width=0.46\textwidth]{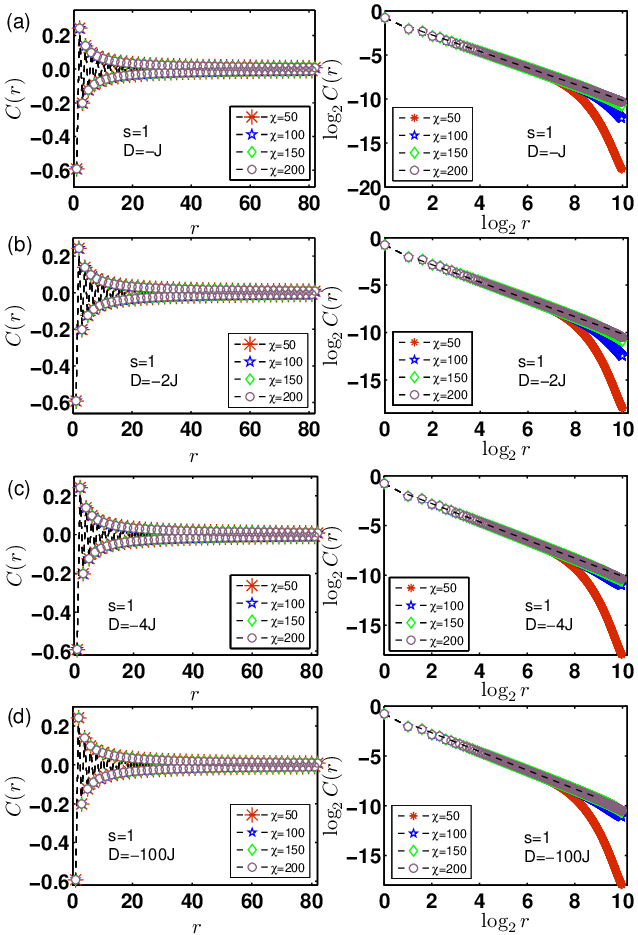}
    \caption{(color online) Left: Spin-spin correlation $C(r)$ as a function of the lattice
    distance $r$. Right: Spin-spin correlation $\log_2 C(r)$ versus lattice distance $\log_2 r$ with $r$ being the even lattice distance, for spin-1 biquadratic XY model.
    Here, we choose four distinct values for (a) $D=-J$, (b) $D=-2J$, (c) $D=-4J$ and
    (d) $D=-100J$ with the truncation dimension
    $\chi=50, 100, 150$ and $200$, respectively. The critical exponent $\eta$ of spin-spin
    correlation is extracted by fitting for the power-law decaying part in (right). Here, we
    take the largest truncation dimension $\chi=200$ to fit.
    }
    \label{Fig24}
\end{figure}
%%%%%%%%%%%%%%%%%%%%%%%Fig. 24%%%%%%%%%%%%%%%%%%%%%%%%%%%%%%%%%%%%%
 Although the magnetization is zero, i.e., $\langle S^{x/y/z}_j\rangle =0$, for the whole parameter range of $D/J$, the behavior of spin correlation can characterize the critical phase. The spin-spin correlation can be defined as ~\cite{Su13}
 \begin{equation}
 C(|i-j|) = \langle S^z_i S^z_j\rangle
 \label{spincorrelation}
 \end{equation}
 with $r=|i-j|$ being the lattice distance. In the left of Fig.~\ref{Fig24}, we plot the spin-spin correlation $C(r)$ as a function of the lattice distance $r$ with various truncation dimensions $\chi=50$, $100$, $150$ and $200$ for (a) $D=-J$, (b) $D= -2J$, (c) $D = -4J$ and $D = -100J$. The spin-spin correlation staggered-oscillates and decays with the increment of lattice distance. The decaying pattern of the spin correlations does not much different for the various chosen values of uniaxial-type single-ion anisotropy.
 In the right of Fig. \ref{Fig24}, the absolute values of the spin-spin correlations are replotted in the log-log scale. It is shown that the absolute values of spin-spin correlations $C(r)$ exhibit a power-law decay as the lattice distance $r$ increases, which means that the ground states are a critical state. Actually, for $D > D_M$, the spin-spin correlations decays exponentially (not presented here), which means that the ground state is not in a critical state. For $\chi=200$, we perform the numerical fitting for the power-law decaying part of the spin correlation in the right of Fig.~\ref{Fig24}.
 With the fitting function $\log_2 C(r)=-\eta \log_2 r + \eta_0$.
 The critical exponents are given as
 (a) $\eta=0.905(5)$ and $\eta_0=-1.01(2)$ for $D=-J$,
 (b) $\eta=0.908(4)$ and $\eta_0=-0.99(2)$ for $D=-2J$,
 (c) $\eta=0.908(4)$ and $\eta_0=-0.99(2)$ for $D=-4J$ and
 (d) $\eta=0.908(4)$ and $\eta_0=-0.99(2)$ for $D=-100J$.
 The estimate critical exponents $\eta$ for various uniaxial-type single-ion anisotropy $D$
 are summarized in Table \ref{table2}. The results show that the estimate exponent does not depend on the uniaxial-type single-ion anisotropy $D$.
 Thus, it is suggested that the estimate critical exponent is to be $\eta \simeq -0.91$ in the critical phase for $\chi=200$.

%%%%%%%%%%%%%%%%TABLE 2%%%%%%%%%%%%%%%%%%%%%
\begin{table}
\renewcommand\arraystretch{2}
\caption{Estimate critical exponents of spin-spin correlation $\eta$ for
 the truncation dimension $\chi=200$ and various values of
 the uniaxial-type single-ion anisotropy $D$ in the critical phase of
 the spin-$1$ system with the rhombic-type single-ion anisotropy $R=0$.
 }
\begin{tabular}{c|cccccc}
\hline
\hline
 $D$ & \begin{minipage}{1.4cm} $-J$  \end{minipage} &
  \begin{minipage}{1.4cm} $-2J$  \end{minipage} &
  \begin{minipage}{1.4cm} $-4J$  \end{minipage} &
  \begin{minipage}{1.4cm} $-100J$  \end{minipage} & \\
 \hline
 \begin{minipage}{1.2cm}  $\eta$ \end{minipage}
 &0.905(5) & 0.908(4) & 0.908(4) & 0.908(4)\\
 \hline
 \hline
\end{tabular}
\label{table2}
\end{table}
%%%%%%%%%%%%%%%TABLE 2%%%%%%%%%%%%%%%%%%%%%%%%%

\subsection{Biaxial spin nematic phases for spin-2 system}
 Similar to the spin-$1$ system in Sec. \ref{Ferro1},
 let us consider the spin-$2$ system with finite uniaxial-type single-ion anisotropy $D$.
 In the limit of very large negative (positive) $R/J \gg D/J$, the spin-$2$ rhombic-type single-ion anisotropy $R$ becomes predominant in the spin-$2$ Hamiltonian (\ref{ham1})
 and thus the local spin state is to be the lowest-energy state of the rhombic-type single-ion anisotropy term
 $R[(S^{x}_j)^2-(S^{y}_j)^2]$ ($R[(S^{y}_j)^2-(S^{x}_j)^2]$),
 i.e.,
 $a|S^x_j=0 \rangle+b|S^y_j=0 \rangle$ ($b|S^x_j=0 \rangle+a|S^y_j=0 \rangle$)
 with $a = (1+\sqrt{3})/\sqrt{6}$ and $b=(1-\sqrt{3})/\sqrt{6}$ \cite{Mao}.
 The ground state becomes the product state of $a|S^x_j=0 \rangle+b|S^y_j=0 \rangle$ ($b|S^x_j=0 \rangle+a|S^y_j=0 \rangle$). Thus, the quadrupole moments are given as
 $\langle Q^{xx}_j\rangle = -\langle Q^{yy}_j \rangle = \sqrt{3}$ and $\langle Q^{zz}_j \rangle=0$ ($\langle Q^{xx}_j\rangle = -\langle Q^{yy}_j \rangle = -\sqrt{3}$ and $\langle Q^{zz}_j \rangle=0$), which implies that both the ground states are a biaxial spin nematic state \cite{Song}. For the limit of very large negative (positive) $R/J \gg D/J$, the spin-$2$ system is in the positive (negative) biaxial spin nematic phase $\langle Q^{x^2-y^2}_j\rangle > 0$ ($\langle Q^{x^2-y^2}_j\rangle < 0$) \cite{Mao}.
 In the following subsections, we will discuss how the transition between the positive and the negative biaxial spin nematic phases occurs as the $R/J$ varies from the negative large value to the positive large value for a given uniaxial-type single-ion anisotropy $D$.

%%%%%%%%%%%%%%%%%%%%%Fig. 25%%%%%%%%%
\begin{figure}
    \includegraphics[width=0.45\textwidth]{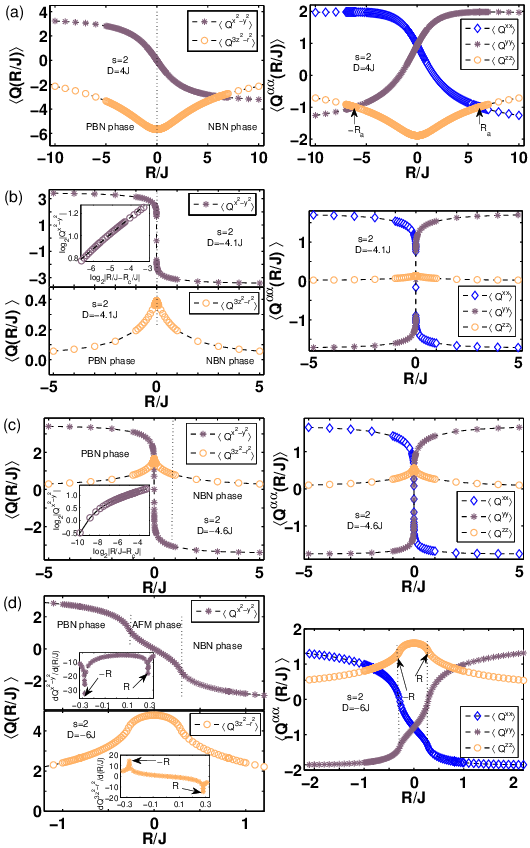}
    \caption{(color online) Quadrupole order parameters $\langle Q_j^{x^2-y^2} \rangle$,
    $\langle Q_j^{3z^2-r^2}\rangle$, and $\langle Q^{\alpha\alpha}\rangle$ ($\alpha \in \{ x,y,z\}$) for (a) $D=4J$, (b) $D=-4.1J$, (c) $D=-4.6J$ and (d) $D=-6J$ with the truncation dimension $\chi=150$ in the spin-$2$ system.
    } \label{Fig25}
     \end{figure}
%%%%%%%%%%%%%%%%%%%%%Fig. 25%%%%%%%%%%%%

\subsubsection{Quantum biaxial spin nematic to nematic crossover and quadrupole moments}
 Let us consider the spin-$2$ system for $D = 4 J$.
 We plot the quadrupole moments $\langle Q^{x^2-y^2}_j \rangle$ and $\langle
 Q^{3z^2-r^2}_j\rangle$ in the left of Fig.~\ref{Fig25}(a) and $\langle Q^{\alpha\alpha}_j \rangle$ $(\alpha=x, y, z)$ in the right of Fig.~\ref{Fig25}(a) as a function of $R/J$ for $\chi=150$. As the $R/J$ increases from the negative large value to the positive large value, i.e., from the positive biaxial spin nematic state to the negative biaxial spin nematic state, the $\langle Q^{x^2-y^2}_j\rangle$ starts to decrease from $2\sqrt{3}$ roughly around $R \sim -5J$, decreases relatively around the inflection point at $R=0$, and saturates to $-2\sqrt{3}$ around $R \sim 5J$. The $\langle Q^{3z^2-r^2}_j\rangle < 0$ shows a monotonous gentle valley shape that has a minimum value at the inflection point $R = 0$.
 The ground state at $R =0$, as shown in Fig. \ref{Fig25}(a), gives  $\langle Q^{xx}_j\rangle = \langle Q^{yy}_j \rangle = -\frac{1}{2}\langle Q^{zz}_j \rangle > 0$
 and thus is to be a uniaxial spin nematic state. At $R=0$, $\langle Q^{x^2-y^2}_j\rangle = 0$, i.e., $\langle Q^{x^2-y^2}_j\rangle >0$ for $R < 0$ and  $\langle Q^{x^2-y^2}_j\rangle <0$ for $R > 0$. However, it is not seen any characteristic behavior of the quadrupole moments indicating a quantum phase transition at the inflection point $R=0$.
 No explicit abrupt phase separation was also observed in the entanglement entropy in Fig. \ref{Fig8}(a) and the WYSIs $I^{x/y/z}$ in Figs. \ref{Fig8}(b) and \ref{Fig8}(c)
 by their second derivatives. Even such similar behaviors of entanglement entropy and WYSIs was found for $D=0$ (not presented here). Furthermore the quadrupole moments for $D=0$ in \cite{Mao} were shown a very similar behavior with those for $D = 4J$ in Fig. \ref{Fig25}(a) without any indication of occurring an explicit phase transition.
 Accordingly, as the $R/J$ varies from $R/J = -\infty$ to $R/J = \infty$,
 the positive and negative biaxial spin nematic states orthogonal to each other
 are connected adiabatically without an explicit abrupt phase transition,
 which is the quantum crossover. For a quantum crossover, a substantial change in the nature of many-body ground state occurs over a finite range of the system parameter rather than abruptly at a critical point \cite{Mao,Khan}. In the biaxial spin nematic phases, the crossover parameter region can be defined as $-R_a \lesssim R_{cross} \lesssim R_a$ with the uniaxial nematic states at $R = \pm R_a$, shown in $\langle Q^{xx/yy/zz}_j \rangle$ in Fig. \ref{Fig25}(a). Such a quantum crossover occurs up to the critical endpoint $D = D_{CEP}$, i.e., for $D > D_{CEP}$.

\subsubsection{Continuous biaxial spin nematic to nematic phase transitions and quadrupole moments}
 Let us consider the uniaxial-type single-ion anisotropy $D$ smaller than $D = D_{CEP}$, e.g., $D = -4.1J$.  The quadrupole moments $\langle Q^{x^2-y^2}_j\rangle$, $\langle Q^{3z^2-r^2}_j\rangle$ and $\langle Q^{\alpha\alpha}_j\rangle$ ($\alpha \in \{ x,y,z\}$)
 are plotted as a function of $R/J$ with $\chi=150$ in Fig.~\ref{Fig25}(b).
 The overall behavior of $\langle Q^{x^2-y^2}_j\rangle$ and $\langle Q^{xx/yy}_j\rangle$ appears to be similar to those in Fig. \ref{Fig25}(a) as $R/J$ increases from negative to positive large values. However, the $\langle Q^{x^2-y^2}_j\rangle$ decreases much sharply in a relatively very narrow range of parameter $R/J$. In contrast to $\langle Q^{3z^2-r^2}_j\rangle < 0$ for the whole parameter range of $R/J$ in Fig. \ref{Fig25}(a),
 $\langle Q^{3z^2-r^2}_j\rangle > 0$ in Fig. \ref{Fig25}(b). Actually, at $D = D_{CEP}$, $\langle Q^{3z^2-r^2}_j\rangle = 0$ and $\langle Q^{xx/yy/zz} \rangle = 0$, i.e., the local spin fluctuates isotropically with $\langle (S^{x/y/z}_j)^2\rangle = 2/3$. Also the $\langle Q^{3z^2-r^2}_j\rangle > 0$ exhibits a cusp shape that has a maximum value at the inflection point $R = 0$.

 As the order parameter for the negative(positive) biaxial spin nematic phase,
 a log-log plot of the $\langle Q^{x^2-y^2}_j\rangle$ near the transition point $R=0$ is displayed in the inset of Fig.~\ref{Fig25}(b). The inset shows a scaling behavior of the quadrupole moment the $\langle Q^{x^2-y^2}_j\rangle$ near the transition point $R=0$.
 Such a behavior of quadrupole moment can be described by the power law, i.e.,
 $\langle Q^{x^2-y^2}_j\rangle \propto |R/J-R_c/J|^\beta$, where $\beta$ is the characteristic critical exponent. To estimate the critical exponent, the least-squares fit of the data near the critical point $R=0$ and thus the estimate critical exponent is given as $\beta=0.094(2)$ for the quadrupole moment. Accordingly, the quadrupole order parameter characterize the continuous biaxial spin nematic to nematic phase transition between the positive (negative) biaxial spin nematic phases. This result is consistent with that of the entanglement entropy and the WYSIs.

%%%%%%%%%%%%%%%%%%%%%%%%%Fig. 26%%%%%%%%%%%%%%%%%%
 \begin{figure}
    \includegraphics[width=0.4\textwidth]{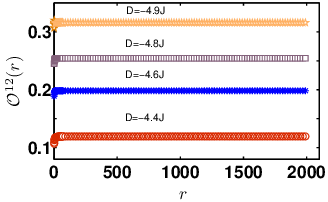}
    \caption{(color online) String order parameter $\mathcal{O}^{12}(r)$ for various
    parameter $D=-4.4J, -4.6J, -4.8J$ and $-4.9J$, as a function of the lattice distance
    $r=|i-j|$, for the spin-2 system on the line $R=0$ with $\chi=150$.
    } \label{Fig26}
     \end{figure}
%%%%%%%%%%%%%%%%%%%%%%%%Fig. 26%%%%%%%%%%%%%%%%%%%%%
\subsubsection{Discontinuous biaxial spin nematic to nematic phase transitions and quadrupole moments}
\label{top}
 As was shown in the entanglement entropy for $\chi=150$ in Fig. \ref{Fig16}(a),
 the ground state of the spin-$2$ system is not critical for $ D_{M} < D < D_{C2}$.
 Then let us consider the uniaxial-type single-ion anisotropy $D = -4.6J$.
 The $\langle Q^{x^2-y^2}_j \rangle$, $\langle Q^{3z^2-r^2}_j\rangle$,
 and $\langle Q^{\alpha\alpha}_j \rangle$ are plotted as a function of $R/J$ in Fig.~\ref{Fig25}(c). Compared with Fig. \ref{Fig25}(b), overall, the behaviors of the quadrupole moments seem to be very much similar to those for $D = -4.1J$ in Fig. \ref{Fig25}(c). Similar to the case of $D = -4.1J$, in order to confirm whether a continuous phase transition occurs or not for $D =-4.6J$, we plot  a log-log plot of the $\langle Q^{x^2-y^2}_j\rangle$ near the transition point $R=0$ in the inset of Fig.~\ref{Fig25}(c).
 The log-log plot shows clearly that the behavior of quadrupole moment near the transition point $R=0$ cannot be described by the power law of $\langle Q^{x^2-y^2}_j\rangle \propto |R/J-R_c/J|^\beta$. Thus the behavior of the quadrupole moment does not indicate an occurrence of continuous phase transition. This result is consistent with that of the entanglement entropy and the WYSIs. 

 As we discussed, the entanglement spectrum at $R = -4.6J$ is doubly degenerate and thus the phase boundary is not a critical state but is a topological state. To characterize the topological phase, we consider the $\rm SO(5)$ symmetry group with the ten generators in the standard spin-$2$ $S^z$ basis $|m\rangle$ ($m = \pm 2, \pm 1,0$) ~\cite{Tu}. The $\rm SO(5)$ Cartan generators are defined as $L^{12} = |2\rangle \langle 2|-|-2\rangle\langle-2|$ and $L^{34}=|1\rangle \langle 1|-|-1\rangle\langle-1|$. In addition, the generators $L^{15}$ and $L^{35}$ are defined as $L^{15} = (|2\rangle |0\rangle + |0\rangle |-2\rangle + \mathrm{h.c.})/\sqrt{2}$ and $L^{35} = (|1\rangle |0\rangle + |0\rangle |-1\rangle +\mathrm{h.c.})/\sqrt{2}$, respectively. Other generators $L^{ab}$ ($1\leq a < b \leq 5$) of the $\rm SO(5)$ symmetry group can be determined by using the $\rm SO(5)$ commutation relation $[L^{ab}, L^{cd}]=-i(\delta_{ac} L^{bd}+ \delta_{bd} L^{ac} - \delta_{ad} L^{bc} - \delta_{bc} L^{ad})$. Accordingly, the topological phase can be characterized by the nonlocal string order parameters of the Catan generators defined as
\begin{equation}
 \mathcal{O}^{12}= \Bigg \langle L_i^{12} \exp \Bigg[i\pi\sum_{k=i+1}^{j-1} L_k^{12} \Bigg ]
 L_{j}^{12} \Bigg \rangle
 \label{String}
\end{equation}
 and $\mathcal{O}^{34}$ replaced $L^{12}$ with $L^{34}$. We have calculated the two string order parameters $\mathcal{O}^{12}$ and $\mathcal{O}^{34}$. Figure \ref{Fig26} displays the string order parameter $\mathcal{O}^{12}(r)$ as a function of the lattice distance $r=|i-j|$ for various uniaxial-type single-ion anisotropies $D=-4.4J$, $-4.6J$, $-4.8J$ and $-4.9J$ with $\chi=150$. The string order parameters $\mathcal{O}^{12}$ tend to saturate to a finite value as the lattice distance $r$ increases. Each string order parameter in Fig. \ref{Fig26} corresponds each spectrum in Fig. \ref{Fig18}. In fact, similar to the string order parameter defined by the spin operator $S^\alpha$, the string order parameter of $L^{34}$ disappears, i.e., $\mathcal{O}^{34}=0$ in the entire parameter region, where the entanglement spectrum is doubly degenerate. As a result, the topological phase is characterized by the two-fold degenerate entanglement spectrum and the nonlocal string order parameter $\mathcal{O}^{12}$
 of the SO(5) Cartan generator $L^{12}$.

\subsubsection{Continuous quantum phase transition from biaxial spin nematic phases
  to spin non-nematic phase and quadrupole moments}
 Let us consider the uniaxial-type single-ion anisotropy $D < D_M$, e.g., $D = -6J$.
 Figure~\ref{Fig25}(d) exhibits the quadrupole moments $\langle Q^{x^2-y^2}_j \rangle$,
 $\langle Q^{3z^2-r^2}_j\rangle$ and $\langle Q^{\alpha\alpha}_j \rangle$ $(\alpha=x, y, z)$
 as a function of $R/J$ for $D = -6J$ with $\chi=150$. Each quadrupole moment for $D = -6J$ in Fig. \ref{Fig25}(d) has two inflection points on both sides within the same parameter distance from $R=0$. One can thus notice that each quadrupole moment for $D = -6J$ in Fig. \ref{Fig25}(d) shows very different behavior compared to those for other parameters, i.e., $D = 4J$, $D = -4.1J$, and $D = -4.6J$ in Figs. \ref{Fig25}(a)-(c), respectively, because the other quadrupole moments have only an inflection point at $R=0$. Based on the properties of the inflection points $R=0$ in Figs. \ref{Fig25}(a) and (b), one may expect that the two inflection points indicate occurring
 continuous or discontinuous phase transitions.

 In fact, near the two inflection points $R = \pm 0.282J$, the quadrupole moments do not reveal any scaling behavior. However, as was shown in the antiferromagnetic order parameter in Fig. \ref{Fig21}(a), the two inflection points of the quadrupole moments correspond to the two transition points, respectively. Accordingly, the change in the direction of curvature of the quadrupole moments, shown in the insets of Fig. \ref{Fig25}(d),
 indicates the continuous biaxial spin nematic to non-nematic phase transition from the positive (negative) biaxial spin nematic phase to the antiferromagnetic phase.

\section{Summary}
\label{summary}
 We have investigated the bipartite entanglement entropy and the WYSI of the ground state in the infinite biquadratic spin-$1$ and spin-$2$ XY chains with the rhombic- and the uniaxial-type single-ion anisotropies by using the iTEBD algorithm in the iMPS representation. The characteristic behaviors of the bipartite entanglement entropy and the WYSIs have been studied concretely in order to catch the various types of quantum phase transitions occurring between the uniaxial spin nematic phases for the spin-$1$ system or the biaxial spin nematic phases for the spin-$2$ system. Indeed, the spin-$1$ and -$2$ systems are described in the same form of Hamiltonian in Eq. (\ref{ham1}), while the spin-$1$ and -$2$ systems show very different features in quantum nematic phases, phase transitions, and phase diagrams, as shown in Figs. \ref{Fig1} and \ref{Fig2}. The bipartite entanglement entropy and the WYSI have been shown to enable to capture various types of quantum spin nematic phase transitions including quantum crossover for the spin-$1$ or spin-$2$ systems.

 For the spin-$1$ system, there are three uniaxial spin nematic quadrupole phases in the whole $(R/J, D/J)$-anisotropy space. By means of their nonanalytical discontinuous behaviors at the transition points, both the bipartite entanglement entropy and the WYIS have captured the discontinuous spin nematic phase transitions occurring from the $z$-ferroquarupole phase to the $x$-ferroquadrupole phase or to the $y$-ferroquadrupole phase. Contrary to the discontinuous uniaxial spin nematic phase transitions, the bipartite entanglement entropy and the firs-order derivatives of WYSIs show a typical singular cusp behavior, and in addition, the cusps become shaper with the increment of the truncation dimension for the continuous uniaxial spin nematic phase transition between the $x$-ferroquadrupole phase and the $y$-ferroquadrupole phase. The central charge at the continuous quantum phase transition point of the spin-$1$ system is estimated as $c \simeq 1$ by using the scaling behavior of the bipartite entanglement entropy. The critical exponent of the spin-spin correlation is also estimated as$\eta \simeq 0.9$ along the phase boundary.

 For the spin-$2$ system, there are two biaxial spin nematic phases and antiferromagnetic phase in the whole $(R/J, D/J)$-anisotropy space. There occur quantum crossover, continuous or discontinuous quantum phase transitions between the two biaxial spin nematic phases for various given uniaxial single-ion anisotropy as the rhombic-type single-ion anisotropy varies. Also, there occurs a continuous phase transition between the antiferromgnetic and the biaxial spin nematic phases. For quantum crossover, both the bipartite entanglement entropy and the WYISs have no explicit nonanalyticcal behavior for the adiabatic connection from one to the other biaxial spin nematic phase. Similar to the continuous uniaxial spin nematic phase transition of the spin-$1$ system, both the bipartite entanglement entropy and the WYSIs show nonanalytical singular cusp behaviors becoming shaper with the increment of the truncation dimension for the continuous biaxial spin nematic phase transition of the spin-$2$ system with the estimate central charge $c \simeq 1$. Contrary to the discontinuous uniaxial spin nematic phase transition of the spin-$1$ system,  both the bipartite entanglement entropy and the WYSIs show a cusp behavior not becoming shaper with the increment of the truncation dimension for the discontinuous biaxial spin nematic phase transition of the spin-$2$ system. At the discontinuous transition point, the ground state is in the topological state identified by the doubly degenerated entanglement spectrum.
 The topological state is specified by the nonlocal string order parameter defined by
 the SO$(5)$ Catan generator $L^{12}$ of the spin-$2$ system. The continuous quantum phase transition from the biaxial spin nematic phase to the antiferromagnetic phase can be detected by the typical nonanalytic cusp behaviors of the bipartite entanglement entropy and the singular behaviors of the firs-order derivatives of the WYSIs. The critical state at the transition points has the central charge $c \simeq 1/2$.

 The local magnetic moments and quadrupole moments have been investigated to determine
 the magnetic and nonmagnetic phases in the spin-$1$ and -$2$ systems.
 For the spin-$1$ system, it is found that there are nonmagnetic nematic phases with the zero local magnetic moments, i.e., $\langle S^\alpha_j\rangle=0$ for the whole $(R/J,D/J)$-anisotropy parameter space. The three nonmagnetic phases are characterized by the local spin quadrupole order parameters as the three distinct uniaxial spin nematic phases, i.e, $x/y/z$-ferroquadrupole phases.
 Similar to the discontinuous behaviors of the bipartite entanglement entropy
 and the WYSIs, the discontinuous quadrupole moments indicate the occurrences of the discontinuous uniaxial spin nematic phase transitions from the $z$-ferroqudarupole phase
 to the $x$- or $y$-ferroquadrupole phases.
 For the continuous uniaxial spin nematic phase transition between the $x$- and the $y$-ferroquadrupole phases, the quarupole order parameter near the transition point
 is shown to be described by  the power law, i.e.,  $\langle Q^{x^2-y^2}_j\rangle \propto |R/J-R_c/J|^\beta$ with $\beta=0.293(3)$.

 In contrast to the spin-$1$ system with the zero local magnetic moment,
 the spin-$2$ system has an antiferromagnetic phase with nonzero local magnetic moments
 as well as the two biaxial spin nematic phases for $D < D_M$. While for $D > D_{M}$, the two biaxial spin nematic phases are separated at the vanishing rhombic-type anisotropy point $R = 0$, according to the sign of the quadrupole order parameter $\langle Q^{x^2-y^2}_j\rangle$.
 There are three types of the phase transition between the two biaxial spin nematic phases at $R =0$. For the quantum crossover, the substantial change of the order parameter arises
 a finite range of the system parameter roughly defined as the range in between
 the two uniaxial nematic states. The two uniaxial nematic states merges into one uniaxial nematic state at $R=0$ as the uniaxial anisotropy approaches to $D=D_{CEP}$. While for the continuous and the discontinuous biaxial spin nematic phase transitions, the quadrupole order parameter $\langle Q^{x^2-y^2}_j\rangle$ reveals a very steep change with an inflection point at $R = 0$  as the rhombic-type single ion anisotropy varies. For the continuous spin nematic phase transition, similar to that of the spin-$1$ system, the quarupole order parameter near the transition point is described by  the power law, i.e.,  $\langle Q^{x^2-y^2}_j\rangle \propto |R/J-R_c/J|^\beta$ with $\beta=0.094(2)$. In contrast to the continuous spin nematic phase transition, the quadrupole order parameter does not follow the power law for discontinuous spin nematic phase transition.

%%%%%%%%%%%%%acknowledgements
\acknowledgements
 This work was supported by
 Technological Innovation 2030- ``Quantum Communication and Quantum Computer" Major
 Project.
 Y.-W.D. acknowledges the support in part from the National Natural Science Foundation of
 China (Grant No. 11805285) and the Research Funds for the Central Universities (Grant No. 2024CDJXY023).

%%%%%%%%%%%%%%%%%%%%%%%%%%%%%%%%%%%%%%%%%%%%

%%%%%%%%%%%%%%%%%%%%%%%%%%%%%%%%%%%%%%EOF%%%%%%%%%%%%%%%

 \end{document}